\documentclass[11pt]{elsarticle}
\usepackage[margin=1.2in]{geometry}
\usepackage{gensymb}
\usepackage{graphicx}
\usepackage{amsmath}
\usepackage{latexsym}
\usepackage{amsfonts}
\usepackage{mathtools}
\usepackage{tcolorbox}
\usepackage{listings}
\usepackage{array}
\usepackage{subcaption}
\usepackage{kantlipsum}
\usepackage{changepage}
\usepackage{dirtytalk}
\usepackage{natbib}
\usepackage[makeroom]{cancel}
\usepackage{amsthm}
\theoremstyle{remark}

\usepackage{epstopdf}

\usepackage{relsize}
\usepackage{float}
\usepackage{pifont}
\usepackage[font={small}]{caption}
\DeclareCaptionFormat{myformat}{\fontsize{9}{10}\selectfont#1#2#3}
\makeatletter
\renewcommand{\fnum@figure}{\textbf{Fig. \thefigure.}}
\makeatother
\makeatletter
\renewcommand{\fnum@table}{\textbf{Tab. \thetable.}}
\makeatother
\usepackage[labelsep=space]{caption}

\usepackage{amssymb}
\usepackage{a4wide}
\usepackage{pst-all}
\usepackage{float}
\usepackage{latexsym}
\usepackage{graphicx}
\usepackage{psfrag}
\usepackage{verbatim}
\usepackage[ruled,vlined]{algorithm2e}
\usepackage{setspace}
\usepackage{xcolor,colortbl}

\renewcommand*\thetable{\Roman{table}}

\newfloat{MyBox}{thp}{loa}
\floatname{MyBox}{Box}

\usepackage{upgreek}

\makeatletter
\def\@author#1{\g@addto@macro\elsauthors{\normalsize%
    \def\baselinestretch{1}%
    \upshape\authorsep#1\unskip\textsuperscript{%
      \ifx\@fnmark\@empty\else\unskip\sep\@fnmark\let\sep=,\fi
      \ifx\@corref\@empty\else\unskip\sep\@corref\let\sep=,\fi
      }%
    \def\authorsep{\unskip,\space}%
    \global\let\@fnmark\@empty
    \global\let\@corref\@empty  
    \global\let\sep\@empty}%
    \@eadauthor={#1}
}
\makeatother

\newcommand{\ds}{\displaystyle}
\newcommand{\bm}[1]{\text{\boldmath $#1$}}

\makeatletter
\def\ps@pprintTitle{%
 \let\@oddhead\@empty
 \let\@evenhead\@empty
 \def\@oddfoot{\centerline{\thepage}}
 \let\@evenfoot\@oddfoot}
\makeatother

\makeatletter
\def\ps@pprintTitle{%
 \let\@oddhead\@empty
 \let\@evenhead\@empty
 \def\@oddfoot{}%
 \let\@evenfoot\@oddfoot}
\makeatother

\begin{document}

\begin{frontmatter}
	\title{On performance bounds for topology optimization}
	\author{Anna Dalklint\fnref{label1}\corref{cor1}}
	\author{Rasmus E. Christiansen \fnref{label1,label2}}
	\author{Ole Sigmund \fnref{label1}}

	\address[label1]{Technical University of Denmark, Department of Civil and Mechanical Engineering, Nils Koppels Allé, Building 404, 2800 Kongens
		Lyngby, Denmark \fnref{label1}}
	\address[label2]{NanoPhoton - Center for Nanophotonics, Technical University of Denmark, Ørsteds Plads 345A, 2800
		Kongens Lyngby, Denmark \fnref{label2}}
	\cortext[cor1]{Corresponding author. E-mail adress: adal@dtu.dk}

\begin{abstract}
Topology optimization has matured to become a powerful engineering design tool that is capable of designing extraordinary structures and materials taking into account various physical phenomena. Despite the method's great advancements in recent years, several unanswered questions remain. This paper takes a step towards answering one of the larger questions, namely: How far from the global optimum is a given topology optimized design? Typically this is a hard question to answer, as almost all interesting topology optimization problems are non-convex. Unfortunately, this non-convexity implies that local minima may plague the design space, resulting in optimizers ending up in suboptimal designs. In this work, we investigate performance bounds for topology optimization via a computational framework that utilizes Lagrange duality theory. This approach provides a viable measure of how \say{close} a given design is to the global optimum for a subset of optimization formulations. The method's capabilities are exemplified via several numerical examples, including the design of mode converters and resonating plates.
\end{abstract}


\end{frontmatter}


\section{Introduction}
Topology optimization has seen significant advancements since its inception in 1988 via the seminal paper by \citet{bendsoe1988generating}. The technique can be used to design remarkably sophisticated structures involving various physical phenomena such as acoustics (\citet{duhring2008acoustic}, \citet{dilgen2024topology}, \citet{christiansen2019topological}), plasticity (\citet{maute1998adaptive}, \citet{wallin2016topology}), contact (\citet{stromberg2010topology}, \citet{bluhm2021internal}), fluids (\citet{borrvall2003topology}, \citet{alexandersen2016large}) and photonics (\citet{jensen2011topology}, \citet{wang2018maximizing}). Via computational and technical developments, the method has even matured to allow the design of high-resolution structures involving billions of design degrees-of-freedom (\citet{aage2017giga}, \citet{liu2018narrow}). For a comprehensive review of the applicability of topology optimization, see \citet{sigmund2013topology}. In this work, we limit ourselves to the density-based approach to topology optimization, although we acknowledge that other approaches exist, cf. \citet{sigmund2013topology}.

Even though the aforementioned applications of topology optimization truly illustrate its power as a design tool, questions remain regarding its efficacy in the sense of design \emph{optimality}. Traditionally, the topology optimization problems are solved in a staggered fashion, i.e. the state variables, e.g. the mechanical displacement or electric field, are given implicitly via the equilibrium equations. This allows the efficient use of gradient-based optimizers such as the method of moving asymptotes (MMA, \citet{svanberg1987method}) by computing the sensitivities using the adjoint method (\citet{tortorelli1994design}). Nonetheless, these optimization problems are rarely convex, wherefore the design space contains local minima that may obstructs the optimizer's search for the optimal design. So-called global optimizers such as Genetic Algorithms, Particle Swarms or Simulated Annealing do not guarantee global optima either, cf. \citet{sigmund2011usefulness}. Historically, the topology optimization community has addressed this issue in various ways. For certain \say{simple} problems, such as compliance minimization, a comparison can be made to state-of-the-art solutions, which can be argued to be at least nearly-optimal from an engineering perspective. Also, certain problems are naturally bounded, such as those involving efficiency metrics, or can be analytically bounded. For those problems, topology optimization has proven to yield near-optimal designs, cf. \citet{christiansen2023inverse}, \citet{sigmund1997design}, \citet{guest2006optimizing}, \citet{andreassen2014design} and \citet{wang2023architecting}. For problems lacking natural or analytical bounds, it is common to resort to heuristic methods wherein the results of a number of optimization runs, each using different initializations of the design variables, are compared. The hope is that the superior design in this set is at least close to the global optimum. Unfortunately, nothing guarantees such a claim.  
In such cases, the topology optimization community would greatly benefit from a measure of how far from optimality a specific design is; in general topology optimization lacks \emph{performance bounds}.

Under certain simplifying assumptions, there can exist analytical bounds to physical problems. One example is the Hashin-Shtrikman bound (\citet{hashin1963variational}), which expresses the limits of isotropic linear elastic composite material properties. Another is the limit of the effective thermal expansion coefficients in relation to the effective elastic moduli of two-phase materials (\citet{rosen1970effective}). In free-space optics, other examples include the optical diffraction limit (\citet{born2013principles}) or bounds on the local density of state (\citet{miller2016fundamental}). Of course, the fundamental bounds of mass, energy and momentum conservation also exist. There has recently been a surge of research that explicitly addresses the development of \emph{computational} bounds. These bounds are, as the name implies, based on numerical simulations, in contrast to analytical or physics-based bounds. In general, the idea behind the computational bounds emanates from traditional Lagrange duality theory; the Lagrange dual function always bounds the primal objective function (\citet{boyd2004convex}). It turns out that the tightest possible bound is found by solving the dual problem to optimality, and this is where the power of the method resides, since the dual problem is guaranteed to be convex even when the primal problem is not.

Of course, the success of the computational bounds relies on the fact that the dual problem can be formulated and solved in an efficient manner, and that its solution does not coincide with the trivial bound. \citet{gertler2023many} argues that many photonic inverse design problems can be posed as quadratically constrained quadratic programs (QCQP), for which the dual (i.e. bound) problem coincides with a convex semidefinite program (SDP) obtained using a so-called semidefinite relaxation technique (\citet{shor1987quadratic}, \citet{luo2010semidefinite}). Indeed, many topology optimization problems involve objectives and constraints that are, or can be reformulated as, quadratic functions in the state field. Unfortunately, the solution times of SDPs typically scale with $n^{3.5}$ or $n^4$ for $n$ degrees-of-freedom, wherefore the application of SDPs to \say{large-scale} optimization problems is still hindered (\citet{gertler2023many}, \citet{luo2010semidefinite}). However, there exists promising recent research that utilizes the sparsity characteristics of the involved matrices to drastically speed up the solution times (\citet{gertler2023many}).

Although the idea of utilizing Lagrange duality to bound optimization problems is far from new, it has only recently been applied to bound inverse design problems. The methodology was exemplified by \citet{angeris2019computational,angeris2021heuristic} to compute bounds for photonic design optimization problems with mode-matching objectives. Following this paper, bounds to various photonics design problems were developed, such as maximum far-field power-bandwidth (\citet{kuang2020computational}), mode converters (\citet{angeris2023bounds}), maximum reflectivity contrast between frequencies (\citet{shim2024fundamental}) and radiative Purcell enhancement (\citet{molesky2020hierarchical}). Recently, the approach was generalized to objectives that are quadratic functions in the field by
\citet{angeris2023bounds} and \citet{gertler2023many}.

Even though the computational bounds have found success in many photonic design problems, their application to other physics remains unexplored. One reason for this is that the so far developed formulations, which target 2D photonics, i.e. scalar-fields, only use Green's function solutions and/or finite difference discretizations of the physics. Many vector-field physical problems, such as elasticity, cannot be accurately and readily solved by finite difference schemes, at least not for arbitrarily complex geometries and boundary conditions. Instead, one usually employs the finite element method in such contexts (\citet{bathe2006finite}). The finite element method is also the traditional approach to topology optimization. This somewhat obstructs the direct translation of the previously developed computational bounds to a general topology optimization context.

In this work, we develop a framework for calculating computational performance bounds in the context of traditional density-based topology optimization. We compute the performance bounds for two different design problems: 1) photonic mode converters and 2) elastic resonating plates, and then solve these problems using the traditional topology optimization approach assessing the achieved performance. Our framework is based on the work by \citet{angeris2023bounds}, which relies on a QCQP reformulation of the primal optimization problem, such that a SDP dual problem is obtained. We exemplify how the framework can be extended to problems wherein the finite element method is employed. Also, in contrast to \citet{angeris2023bounds}, we introduce additional cross-correlation constraints in the formulation which tightens the bounds.

\section{Preliminaries}
We seek performance bounds to non-convex optimization problems on the form
\begin{equation}
	\begin{cases}
		\begin{array}{l}
			\underset{\bm{z},\bm{\theta}}{\text{max}} \ \ f(\bm{z}) \\[10pt]
			\text{s.t.} \ \ \begin{cases}
			\bm{A}(\bm{\theta}) \bm{z} = \bm{b}, \\[5pt]
			\bm{\theta} \in \mathcal{B},
			 \end{cases}
		\end{array}
	\end{cases}
	\label{opt1}
\end{equation}
where the set $\mathcal{B}$ defines the box constraints,
$\bm{z}\in\mathbb{C}^n$ is the \emph{finite element} discretized complex state field, whereas $\bm{\theta}\in\mathbb{R}^n$ is the real discretized design density field, where $n$ is the number of degrees-of-freedom. The equality constraint in \eqref{opt1} stems from the PDE that $\bm{z}$ must fulfill (for e.g. elasticity $\bm{A}$ is the stiffness matrix, $\bm{z}$ the displacement field and $\bm{b}$ is the external force), where $\bm{A}(\bm{\theta})\in\mathbb{C}^{n\times n}$ is Hermitian and $\bm{b} 	\in\mathbb{C}^{n}$. We follow \citet{angeris2023bounds} and further limit \eqref{opt1} to objectives which are ratios of \emph{quadratic} functions, i.e.
\begin{equation}
	f(\bm{z}) = \frac{\bar{\bm{z}}\bm{P}\bm{z} + 2\Re(\bar{\bm{p}}\bm{z}) + r}{\bar{\bm{z}}\bm{Q}\bm{z} + 2\Re(\bar{\bm{q}}\bm{z}) + s},
	\label{obj}
\end{equation}
where $\bar{(\,\cdot \,)}$ denotes the conjugate transpose,
$\bm{P},\bm{Q}\in\mathbb{R}^{n\times n}$ are real symmetric matrices,
$\bm{p},\bm{q}\in\mathbb{C}^{n}$ and $r,s\in\mathbb{R}$.

Existing frameworks for performance bounds of inverse problems (e.g. \citet{angeris2023bounds}) relies on certain characteristics of the optimization problem. This explains why \eqref{opt1} is somewhat simplified in comparison to a conventional topology optimization problem; only the PDE and box constraints are included, no additional constraints such as a volume constraint is present. As mentioned in the introduction, the techniques utilized herein to compute the performance bounds rely on a QCQP reformulation of \eqref{opt1}, wherefore the incorporation of arbitrary constraints is not straightforward, as will be obvious later on. Nonetheless, ideas of how to include a volume constraint exist (\citet{kuang2023fundamental}). However, in our numerical examples, the volume constraint is not explicitly needed, wherefore we can simplify the formulation by excluding it. Incorporation of more complex constraints, such as stress constraints or eigenfrequency constraints, still remains an unexplored avenue.

Certain properties are also required for the equilibrium constraint $\bm{A}(\bm{\theta})\bm{z} = \bm{b}$ for it to conform to the performance bound framework. Firstly, it should be possible to split the matrix $\bm{A}$ into two matrices; one independent of $\bm{\theta}$ and one dependent of $\bm{\theta}$, such that $\bm{A}(\bm{\theta}) = \bm{C} + \bm{D}(\bm{\theta})$. Here it is worth emphasizing that $\bm{C}=\bm{0}$ is a valid choice. Secondly, the design dependency should enter $\bm{D}(\bm{\theta})$ in a specific way, namely $\bm{D}(\bm{\theta}) = \text{diag}(\bm{\theta}) \text{diag}(\bm{d})$. We acknowledge that these restrictions are rather harsh, and strictly limit the applicability of the methodology to a subset of design problems. In a traditional finite element setting, the aforementioned conditions basically reduces the design problem space to dynamic problems where the design density variables $\bm{\theta}$ dictates the mass distribution, since the mass matrix can be made diagonal using lumping techniques, such that $\bm{\theta}$ distributes point masses in the degrees-of-freedom. There exists ideas of how to extend the method to problems where the above conditions do not hold, i.e. static problems (cf. \citet{angeris2022note}), but we leave the exploration of such problems to future work.

\subsection{Real field reformulation}
In the following, we base the derivations on the aforementioned assumptions. The first task is to reformulate the complex-field equilibrium constraint in \eqref{opt1} into its real-field equivalent, which allows the use of traditional optimization solvers. To this end, a usual complex split can be utilized such that
\begin{equation}
	\bm{A}(\bm{\theta}) \bm{z} = \bm{b} \ \ \Leftrightarrow \ \
	\begin{bmatrix}
		\Re(\bm{A}(\bm{\theta})) & - \Im(\bm{A}(\bm{\theta})) \\[10pt]
		\Im(\bm{A}(\bm{\theta})) & \Re(\bm{A}(\bm{\theta}))\end{bmatrix} \begin{bmatrix}
		\Re(\bm{y}) \\[10pt]
		\Im(\bm{y})
	\end{bmatrix} = \begin{bmatrix}
		\Re(\bm{b}), \\[10pt]
		\Im(\bm{b})\end{bmatrix},
	\label{expandEqui1}
\end{equation}
where $\Re(\, \cdot \,)$ and $\Im(\, \cdot \,)$ denote the real and imaginary part operations, respectively. By inserting the assumption $\bm{A}(\bm{\theta}) = \bm{C} + \bm{D}(\bm{\theta})$ in \eqref{expandEqui1}, we obtain
\begin{equation}
	\left(\begin{bmatrix}
			\Re(\bm{C}) & - \Im(\bm{C}) \\[10pt]
			\Im(\bm{C}) & \Re(\bm{C})
		\end{bmatrix} +
	\begin{bmatrix}
			\text{diag}(\bm{\theta}) & \bm{0}                   \\[10pt]
			\bm{0}                   & \text{diag}(\bm{\theta})
		\end{bmatrix}
	\begin{bmatrix}
			\Re(\text{diag}(\bm{d})) & - \Im(\text{diag}(\bm{d})) \\[10pt]
			\Im(\text{diag}(\bm{d})) & \Re(\text{diag}(\bm{d}))\end{bmatrix} \right)
	\begin{bmatrix}
		\Re(\bm{z}) \\[10pt]
		\Im(\bm{z})
	\end{bmatrix} = \begin{bmatrix}
		\Re(\bm{b}) \\[10pt]
		\Im(\bm{b})\end{bmatrix},
	\label{expandEqui2}
\end{equation}
which is denoted
\begin{equation}
	\left(\bm{C}' + \text{diag}(\bm{\theta}')\bm{D}'\right)\bm{z}' = \bm{b}'.
	\label{expandEqui3}
\end{equation}
Here it is noted that $\bar{\theta}'\in\mathbb{R}^{2n}$, i.e. the number of density design variables has doubled to account for the complex-valued nature of the original problem. To retain the original design parametrization, the additional constraints
\begin{equation}
	\theta'_{n+j} = \theta'_j, \quad j=1,2,...,n,
	\label{thetaConst1}
\end{equation}
must be appended to \eqref{opt1}. The optimization problem in \eqref{opt1} is to this end restated as
\begin{equation}
	\begin{cases}
		\begin{array}{l}
			\underset{\bm{z}',\bm{\theta}'}{\text{max}} \ \ \ds\frac{\bar{\bm{z}}\bm{P}\bm{z} + 2\Re(\bar{\bm{p}}\bm{z}) + r}{\bar{\bm{z}}\bm{Q}\bm{z} + 2\Re(\bar{\bm{q}}\bm{z}) + s} \\[15pt]
			\text{s.t.} \ \ \begin{cases}
			\left(\bm{C}' + \text{diag}(\bm{\theta}')\bm{D}'\right)\bm{z}' = \bm{b}', \\[4pt]
			\theta'_{n+j} = \theta'_j, \quad j=1,2,...,n,                             \\[5pt]
			{\theta}_j' \in [-1,1], \quad j=1,2,...,n.
			\end{cases}
		\end{array}
	\end{cases}
	\label{opt2}
\end{equation}
It is emphasized that the so-called cross-correlation constraints in
\eqref{thetaConst1} are neglected by \citet{angeris2023bounds}. Based on our numerical investigations, we have found that these constraints are of critical importance to obtain useful bounds. Of course, in cases where the physics can be described by real fields, the necessity of these constraints vanish.

\section{Topology optimization}
In our topology optimization, we solve \eqref{opt1} staggered, i.e. introduce the implicit dependency $\bm{z}(\bm{\theta}) = \bm{A}(\bm{\theta})^{-1}\bm{b}$, such that \eqref{expandEqui1} is solved explicitly in each design iteration. The topology optimization problem the optimizer sees therefore reads
\begin{equation}
	\begin{cases}
		\begin{array}{l}
			\underset{\bm{\theta}}{\text{min}} \ \ f(\bm{z}(\bm{\theta})), \\[8pt]
			\text{s.t.} \ \
			\bm{\theta} \in \mathcal{B}.
		\end{array}
	\end{cases}
	\label{Topopt1}
\end{equation}
Throughout this paper, it is assumed that the box constraints are defined as $\mathcal{B} = \{\bm{\theta}\in\mathbb{R}^n: \theta_j \in [-1,1], \ j = 1,2,...,n\}$. We emphasize that we bound the density design variables between $[-1,1]$ for numerical reasons that will become apparent when deriving the bounds. For convenience, we further assume that the design variables $\bm{\theta}$ scale a material property (e.g. the relative permittivity or the mass density) linearly between two values $a$ and $a_o$, i.e.
\begin{equation}
	a \leftarrow \frac{1}{2}\left(\theta_j\left(a-a_o\right) +  \left(a+a_o\right)\right), \ \ j = 1,...,d,
	\label{densScale}
\end{equation}
although other choices as valid as we will discuss later on.

The topology optimization problem in \eqref{Topopt1} is solved using the  gradient-based optimizer MMA (\citet{svanberg1987method}). To this end, the sensitivities of the objective function with respect to $\bm{\theta}$ are necessary. Via the adjoint method, we find
\begin{equation}
	\frac{\partial f(\bm{z}(\bm{\theta}))}{\partial \bm{\theta}} = - \begin{bmatrix}
		\Re(\bm{\mu})^T & \Im(\bm{\mu})^T
	\end{bmatrix} \begin{bmatrix}
		\frac{\partial \Re(\bm{A})}{\partial \bm{\theta}} & -\frac{\partial \Im(\bm{A})}{\partial \bm{\theta}} \\[10pt]
		\frac{\partial \Im(\bm{A})}{\partial \bm{\theta}} & \frac{\partial \Re(\bm{A})}{\partial \bm{\theta}}
	\end{bmatrix} \begin{bmatrix}
		\Re(\bm{z}) \\[10pt]
		\Im(\bm{z})
	\end{bmatrix},
	\label{sens1}
\end{equation}
where $\Re(\bm{\mu}),\Im(\bm{\mu})\in\mathbb{R}^n$ are the adjoint variables obtained from the solution of the linear system
\begin{equation}
	\begin{bmatrix}
		\Re(\bm{A})  & \Im(\bm{A}) \\[10pt]
		-\Im(\bm{A}) & \Re(\bm{A})
	\end{bmatrix}
	\begin{bmatrix}
		\Re(\bm{\mu}) \\[10pt]
		\Im(\bm{\mu})
	\end{bmatrix} =
	\begin{bmatrix}
		\frac{\partial f}{\partial \Re(\bm{z})} \\[10pt]
		\frac{\partial f}{\partial \Im(\bm{z})}
	\end{bmatrix},
	\label{sens2}
\end{equation}
where
\begin{equation}
	\begin{array}{l}
		\ds \frac{\partial f}{\partial \Re(\bm{z})} = \frac{2\bm{P}\Re(\bm{z}) + 2\Re(\bm{p})}{\bar{\bm{z}}\bm{Q}\bm{z} + 2\Re(\bar{\bm{q}}\bm{z}) + s} - \left(2\bm{Q}\Re(\bm{z}) + 2\Re(\bm{q})\right)\frac{\bar{\bm{z}}\bm{P}\bm{z} + 2\Re(\bar{\bm{p}}\bm{z}) + r}{\left(\bar{\bm{z}}\bm{Q}\bm{z} + 2\Re(\bar{\bm{q}}\bm{z}) + s\right)^2}, \\[15pt]
		\ds \frac{\partial f}{\partial \Im(\bm{z})} = \frac{2\bm{P}\Im(\bm{z}) + 2\Im(\bm{p})}{\bar{\bm{z}}\bm{Q}\bm{z} + 2\Re(\bar{\bm{q}}\bm{z}) + s} - \left(2\bm{Q}\Im(\bm{z}) + 2\Im(\bm{q})\right)\frac{\bar{\bm{z}}\bm{P}\bm{z} + 2\Re(\bar{\bm{p}}\bm{z}) + r}{\left(\bar{\bm{z}}\bm{Q}\bm{z} + 2\Re(\bar{\bm{q}}\bm{z}) + s\right)^2}.
	\end{array}
	\label{sens3}
\end{equation}

\subsection{Differences to a conventional topology optimization approach}
As already discussed, our topology optimization problem in \eqref{Topopt1} does in some ways differ from a conventional topology optimization formulation. Firstly, no penalization scheme is used when interpolating the material properties in \eqref{densScale}. Of course, it is acknowledged that penalization schemes are frequently incorporated in a topology optimization context, with popular choices including SIMP (\citet{bendsoe1989optimal}) and RAMP (\citet{stolpe2001alternative}). Nothing hinders the introduction of such a scheme in our formulation, but to keep the presentation as clean as possible we refrain from using any penalization. Lastly, we do not use any filtering technique. This is the case since the introduction of a filter inherently results in a loss of spatial locality of the design variables, as they become functions of their neighbors. This complicates the computation of the bounds, wherefore we leave the inclusion of filtering in the bounds to future work. Nonetheless, since the design space of a filtered topology optimization problem is a subset of the design space of the original topology optimization problem in \eqref{Topopt1}, it is emphasized that bounds to \eqref{Topopt1} are still valid as non-tight bounds to a filtered topology optimization problem. We will utilize this property in the numerical results.

\section{Performance bounds}
The derivation of our performance bounds follows closely those of \citet{angeris2023bounds}. For clarity, we repeat most of those derivations here. 

\subsection{Reformulation of fractional objective}
The first task is to reformulate our objective function in \eqref{obj} such that it is not a fraction. The reason why this is done will become apparent later; it is to obtain an optimization problem on standard QCQP form. To this end, \eqref{expandEqui3} is multiplied by the parameter $\alpha\in\mathbb{R}$, such that
\begin{equation}
	\left(\bm{C}' + \text{diag}(\bm{\theta}')\bm{D}'\right)\bm{y}' = \alpha\bm{b}',
	\label{eqScale}
\end{equation}
where $\bm{y}' = \alpha\bm{z}'$ (or equivalently $\bm{y} = \alpha\bm{z}$). By plugging in $\bm{z} = \bm{y}/\alpha$ in \eqref{obj} we find
\begin{equation}
	f(\bm{y}) = \frac{\bar{\bm{y}}\bm{P}\bm{y} + 2\alpha\Re(\bar{\bm{p}}\bm{y}) + \alpha^2 r}{\bar{\bm{y}}\bm{Q}\bm{y} + 2\alpha\Re(\bar{\bm{q}}\bm{y}) + \alpha^2 s}.
	\label{g011}
\end{equation}
The idea is now to let $\alpha$ scale the input excitation $\bm{b}$, such that the denominator of \eqref{g011} equals one. To this end, $\alpha$ is introduced to \eqref{opt2} as an auxiliary design variable. The optimization problem together with the equality constraint becomes
\begin{equation}
	(\mathbb{OPT})\begin{cases}
		\begin{array}{l}
			\underset{\bm{y}',\bm{\theta}',\alpha}{\text{max}} \ \ \bar{\bm{y}}\bm{P}\bm{y} + 2\alpha\Re(\bar{\bm{p}}\bm{y}) + \alpha^2 r \\[10pt]
			\text{s.t.} \ \ \begin{cases}
			\bar{\bm{y}}\bm{Q}\bm{y} + 2\alpha\Re(\bar{\bm{q}}\bm{y}) + \alpha^2 s = 1, \\[5pt]
			\left(\bm{C}' + \text{diag}(\bm{\theta}')\bm{D}'\right)\bm{z}' = \bm{b}',   \\[4pt]
			\theta'_{n+j} = \theta'_j, \quad j=1,2,...,n,                               \\[5pt]
			{\theta}_j' \in [-1,1], \quad j=1,2,...,n.
			\end{cases}
		\end{array}
	\end{cases}
	\label{optHomo}
\end{equation}
which can be shown to be equivalent to \eqref{opt2} if $\bm{A}(\bm{\theta})$ is invertible, cf. \citet{angeris2023bounds}.

\subsection{Variable elimination}
Next, we show that it is possible to eliminate the density design variables $\bm{\theta}$ from the optimization problem, i.e. obtain an optimization problem that solely uses $\bm{y}'$ and $\alpha$ as design variables. To this end, \eqref{expandEqui3} is rearranged and rewritten on component-wise form such that
\begin{equation}
	\bm{C}_{j,*}'\bm{y}' - \alpha b_j = -\theta_j'\bm{D}_{j,*}'\bm{y}', \quad j=1,2,...,2n,
	\label{thetaElim3}
\end{equation}
where e.g. $\bm{C}_{j,*}'$ denotes the $j$th row of $\bm{C}'$. By taking the absolute value on both sides of the above, it can be identified that
\begin{equation}
	\vert \bm{C}_{j,*}'\bm{y}' - \alpha b_j\vert = \vert -\theta_j'\bm{D}_{j,*}'\bm{y}' \vert  = \vert \theta_j'\vert \vert\bm{D}_{j,*}'\bm{y}' \vert \leq \vert \bm{D}_{j,*}\bm{y}'\vert, \quad j=1,2,...,2n,
	\label{thetaElim4}
\end{equation}
since $\theta_j'\in[-1,1]$ yields $\vert\theta_j'\vert \leq 1$. For later convenience, both sides of \eqref{thetaElim4} are squared, which renders
\begin{equation}
	\vert \bm{C}_{j,*}'\bm{y}' - \alpha b_j\vert^2 \leq \vert \bm{D}_{j,*}'\bm{y}'\vert^2, \quad j=1,2,...,2n.
	\label{thetaElim5}
\end{equation}
The field $\bm{y}'$ satisfies \eqref{eqScale} for a given $\bm{\theta}'$ if and only if it satisfies the above inequality constraints in \eqref{thetaElim5}, cf. \citet{angeris2023bounds}. Therefore, by enforcing the above constraints in our optimization problem, $\bm{\theta}$ is ultimately eliminated.

It is possible to compress the number of constraints in \eqref{thetaElim5}. Inserting the definitions of the matrices/vectors in \eqref{expandEqui2} in \eqref{thetaElim5}, we obtain two constraints for each degree-of-freedom $j=1,2,...,n$, i.e.
\begin{equation}
	\begin{array}{ll}
		\vert \Re(\bm{C}_{j,*})\Re(\bm{y}) - \Im(\bm{C}_{j,*})\Im(\bm{y}) -\alpha\Re({b}_j)\vert^2 \leq \vert\Im(\bm{D}_{j,*})\Im(\bm{y}) - \Re(\bm{D}_{j,*})\Re(\bm{y})\vert^2, \\[10pt]
		\vert\Im(\bm{C}_{j,*})\Re(\bm{y}) + \Re(\bm{C}_{j,*})\Im(\bm{y}) -\alpha\Im({b}_j)\vert^2 \leq  \vert\Im(\bm{D}_{j,*})\Re(\bm{y}) + \Re(\bm{D}_{j,*})\Im(\bm{y})\vert^2.
	\end{array}
	\label{thetaElim5exp}
\end{equation}
By combining the above constraints for each $j = 1,2,...,n$, expanding the squares and introducing $\bm{x} = [\Re(\bm{y}), \, \Im(\bm{y}), \, \alpha]^T$, the quadratic constraints
\begin{equation}
	\bm{x}^T{\bm{A}}'_j\bm{x} \leq 0, \ \ j = 1,2,...,n,
	\label{thetaElim6}
\end{equation}
are obtained, where
\begin{equation}
	{\bm{A}}'_j = \begin{bmatrix}
		\bm{A}_{11}   & \bm{A}_{12}   & \bm{A}_{13} \\[10pt]
		\bm{A}_{12}^T & \bm{A}_{11}   & \bm{A}_{23} \\[10pt]
		\bm{A}_{13}^T & \bm{A}_{23}^T & \bm{A}_{33}
	\end{bmatrix},
	\label{thetaElim7}
\end{equation}
and
\begin{equation}
	\begin{array}{ll}
		\bm{A}_{11} = \Re(\bm{C}_{j,*})^T\Re(\bm{C}_{j,*}) + \Im(\bm{C}_{j,*})^T\Im(\bm{C}_{j,*}) -  \Re(\bm{D}_{j,*})^T\Re(\bm{D}_{j,*})- \Im(\bm{D}_{j,*})^T\Im(\bm{D}_{j,*}), \\[10pt]
		\bm{A}_{12} = -\Re(\bm{C}_{j,*})^T\Im(\bm{C}_{j,*}) + \Im(\bm{C}_{j,*})^T\Re(\bm{C}_{j,*}) + \Re(\bm{D}_{j,*})^T\Im(\bm{D}_{j,*})- \Im(\bm{D}_{j,*})^T\Re(\bm{D}_{j,*}), \\[10pt]
		\bm{A}_{13} = -\Re(b_j)\Re(\bm{C}_{j,*})^T -\Im(b_j)\Im(\bm{C}_{j,*})^T ,                                                                                                \\[10pt]
		\bm{A}_{23} = \Re(b_j)\Im(\bm{C}_{j,*})^T -\Im(b_j)\Re(\bm{C}_{j,*})^T,                                                                                                  \\[10pt]
		\bm{A}_{33} = \Re(b_j)^2 + \Im(b_j)^2.
	\end{array}
	\label{As}
\end{equation}

\subsubsection{Cross-correlation constraints}
The next task is to reformulate the constraints in \eqref{thetaConst1} such that the explicit dependency of $\bm{\theta}$ is avoided. By rearranging \eqref{thetaElim3}, the relationship
\begin{equation}
	\theta_j' = \frac{\alpha b_j - \bm{C}_{j,*}'\bm{y}'}{\bm{D}_{j,*}'\bm{y}'}, \quad j = 1,2,...,2n,
	\label{thetaConst2}
\end{equation}
is found. It is emphasized that \eqref{thetaConst2} is well-defined as long as $\bm{D}_{j,*}'\bm{y}'\neq 0$, which should never happen in practice. If $\bm{D}_{j,*}'\bm{y}'= 0$, the associated $\theta_j$ does not affect the analysis, i.e. this $\theta_j$ is ill-defined in the first case. Inserting the above in \eqref{thetaConst1} renders
\begin{equation}
	\left(\alpha b_{n+j} - \bm{C}_{n+j,*}'\bm{y}'\right)\bm{D}_{j,*}'\bm{y}' = \left(\alpha b_{j} - \bm{C}_{j,*}'\bm{y}'\right)\bm{D}_{n+j,:}'\bm{y}', \quad j = 1,2,...,n,
	\label{thetaConst3}
\end{equation}
which is equivalent to
\begin{equation}
	\begin{array}{ll}
		\left(\alpha \Im(b_{j}) - \Im(\bm{C}_{j,*})\Re(\bm{y}) - \Re(\bm{C}_{j,*})\Im(\bm{y})\right)\left(\Re(\bm{D}_{j,*})\Re(\bm{y}) - \Im(\bm{D}_{j,*})\Im(\bm{y})\right) = \\[10pt]
		\left(\alpha \Re(b_{j}) - \Re(\bm{C}_{j,*})\Re(\bm{y}) + \Im(\bm{C}_{j,*})\Im(\bm{y})\right)\left(\Im(\bm{D}_{j,*})\Re(\bm{y}) + \Re(\bm{D}_{j,*})\Im(\bm{y})\right), \quad j = 1,2,...,n,
	\end{array}
	\label{thetaConst4}
\end{equation}
where \eqref{expandEqui2} was used. Similarly to \eqref{thetaElim5exp}, we can rewrite \eqref{thetaConst4} on the quadratic format
\begin{equation}
	\bm{x}^T{\bm{B}}'_j\bm{x} = 0,
	\label{thetaConst5}
\end{equation}
where
\begin{equation}
	{\bm{B}}'_j = \begin{bmatrix}
		\bm{B}_{11} & \bm{B}_{12} & \bm{B}_{13} \\[10pt]
		\bm{B}_{21} & \bm{B}_{11} & \bm{B}_{23} \\[10pt]
		\bm{0}      & \bm{0}      & 0
	\end{bmatrix},
	\label{thetaConst6}
\end{equation}
and
\begin{equation}
	\begin{array}{ll}
		\bm{B}_{11} = -\Im(\bm{C}_{j,*})\Re(\bm{D}_{j,*})^T+\Re(\bm{C}_{j,*})\Im(\bm{D}_{j,*})^T \\[10pt]
		\bm{B}_{12} = \Im(\bm{C}_{j,*})\Im(\bm{D}_{j,*})^T+\Re(\bm{C}_{j,*})\Re(\bm{D}_{j,*})^T  \\[10pt]
		\bm{B}_{13} =  \Im(b_j)\Re(\bm{D}_{j,*})-\Re(b_j)\Im(\bm{D}_{j,*})                       \\[10pt]
		\bm{B}_{21} = -\Re(\bm{C}_{j,*})\Re(\bm{D}_{j,*})^T-\Im(\bm{C}_{j,*})\Im(\bm{D}_{j,*})^T \\[10pt]
		\bm{B}_{23} =  -\Im(b_j)\Im(\bm{D}_{j,*})-\Re(b_j)\Re(\bm{D}_{j,*})
	\end{array}.
	\label{thetaConst7}
\end{equation}

\subsection{The QCQP}
All the above derivations allows us to rewrite \eqref{optHomo} to the form
\begin{equation}
	\begin{cases}
		\begin{array}{l}
			\underset{\bm{x}}{\text{max}} \ \  \bm{x}^T{\bm{P}}'\bm{x}, \\[5pt]
			\text{s.t.} \ \ \begin{cases}
			\bm{x}^T{\bm{Q}}'\bm{x} = 1,                        \\[5pt]
			\bm{x}^T{\bm{A}}'_j \bm{x} \leq 0, \ \ j = 1,...,n, \\[5pt]
			\bm{x}^T{\bm{B}}'_j \bm{x} = 0, \ \ j = 1,...,n,
			\end{cases}
		\end{array}
	\end{cases}
	\label{optElim}
\end{equation}
where we introduced
\begin{equation}
	{\bm{P}}' = \left[\begin{matrix}
			\bm{P}        & \bm{0}        & \Re(\bm{p}) \\[5pt]
			\bm{0}        & \bm{P}        & \Im(\bm{p}) \\[5pt]
			\Re(\bm{p})^T & \Im(\bm{p})^T & r
		\end{matrix}\right] \ \ \text{ and } \ \ {\bm{Q}}' = \left[\begin{matrix}
			\bm{\bm{Q}}   & \bm{0}        & \Re(\bm{q}) \\[5pt]
			\bm{0}        & \bm{Q}        & \Im(\bm{q}) \\[5pt]
			\Re(\bm{q})^T & \Im(\bm{q})^T & s
		\end{matrix}\right],
	\label{g0P}
\end{equation}
such that
${\bm{P}}',{\bm{Q}}',{\bm{A}}'_j,{\bm{B}}'_j\in\mathbb{R}^{2n+1,2n+1}$ and $\bm{x}\in\mathbb{R}^{2n+1}$. The above optimization problem in \eqref{optElim} is a QCQP, and we note that it is convex if and only if ${\bm{P}}'$, ${\bm{Q}}'$, ${\bm{A}}'_j$ and ${\bm{B}}'_j$, $j=1,...,n$ are positive semidefinite matrices, i.e. in general it is non-convex.

\subsection{Semidefinite relaxation}
The optimization problem in \eqref{optElim} can be rewritten to a SDP using the trick that
\begin{equation}
	\bm{x}^T{\bm{P}}'\bm{x} = \text{trace}\left(\bm{x}^T{\bm{P}}'\bm{x}\right) = \text{trace}\left({\bm{P}}'\bm{x}\bm{x}^T\right) = \text{trace}\left({\bm{P}}'\bm{X}\right),
	\label{trick}
\end{equation}
where $\bm{X} = \bm{x}\bm{x}^T$ is a rank one symmetric positive semidefinite matrix. The QCQP in \eqref{optElim} is therefore equivalent to the SDP
\begin{equation}
	\begin{cases}
		\begin{array}{l}
			\underset{\bm{X}}{\text{max}} \ \  \text{trace}\left({\bm{P}}'\bm{X}\right), \\[5pt]
			\text{s.t.} \ \ \begin{cases}
				                \text{trace}\left({\bm{Q}}'\bm{X}\right) = 1,                      \\[5pt]
				                \text{trace}\left({\bm{A}}'_j\bm{X}\right)\leq 0, \ \ j = 1,...,n, \\[5pt]
				                \text{trace}\left({\bm{B}}'_j\bm{X}\right)= 0, \ \ j = 1,...,n,    \\[5pt]
				                \text{rank}(\bm{X}) = 1,                                           \\[5pt]
				                \bm{X} \succeq 0,
			                \end{cases}
		\end{array}
	\end{cases}
	\label{optSDP}
\end{equation}
where $\bm{X} \succeq 0$ constrains $\bm{X}$ to the set of symmetric positive semidefinite matrices. In general, the SDP in \eqref{optSDP} is a non-convex optimization problem due to the rank-one constraint on $\bm{X}$, cf. \citet{luo2010semidefinite}. However, by simply dropping this constraint, i.e. utilizing the semidefinite relaxation technique (\citet{shor1987quadratic}), we obtain the convex optimization problem
\begin{equation}
	\begin{cases}
		\begin{array}{l}
			\underset{\bm{X}}{\text{max}} \ \  \text{trace}\left({\bm{P}}'\bm{X}\right), \\[5pt]
			\text{s.t.} \ \ \begin{cases}
				                \text{trace}\left({\bm{Q}}'\bm{X}\right) = 1,                      \\[5pt]
				                \text{trace}\left({\bm{A}}'_j\bm{X}\right)\leq 0, \ \ j = 1,...,n, \\[5pt]
				                \text{trace}\left({\bm{B}}'_j\bm{X}\right)= 0, \ \ j = 1,...,n,    \\[5pt]
				                \bm{X} \succeq 0.
			                \end{cases}
		\end{array}
	\end{cases}
	\label{optRelaxSDP}
\end{equation}
The solution to \eqref{optRelaxSDP} will always bound the solution of \eqref{opt2} from above, since it can be shown to be equivalent to the dual problem of \eqref{optElim}, cf. \citet{vandenberghe1996semidefinite}.

While we via the aforementioned steps have gained a convex optimization
problem, we have lost computational efficiency. The number of design variables increased from $n$ to $n^2$, which causes not only an increase in computational time to solve \eqref{optRelaxSDP}, but also a drastic increase in the memory demand since $\bm{X}$ is dense. It is however emphasized that the solution to \eqref{optRelaxSDP} can be obtained more efficiently by solving the dual problem, which we do, see \citet{angeris2023bounds} for more details.

\subsection{What information does the bounds provide?}
The solution to \eqref{optRelaxSDP} bounds the solution to the primal problem in \eqref{opt2}. However, when solving \eqref{optRelaxSDP}, we also obtain the solution variables $\bm{X}$. The question is if and how we can relate $\bm{X}$ to $\bm{x}$, and thereby the physical field $\bm{z}$ and density design field $\bm{\theta}$. If a clearly interpretable relationship can be identified, the bound problem could be used to not only bound the primal problem objective, but also to generate designs that might prove useful as e.g. initial guesses to the topology optimization problem, cf. \citet{angeris2019computational}, and expanded on in the following.
 
Indeed, the transformation of the QCQP in \eqref{optElim} to the SDP
in \eqref{optSDP} includes the explicit definition $\bm{X} = \bm{x}\bm{x}^T$. However, while the relaxation gave us a convex optimization problem, it also resulted in a loss of this exact relation between $\bm{X}$ and $\bm{x}$, since the rank-1 constraint was neglected. In general, the design variable matrix $\bm{X}$ will instead have a rank$(\bm{X})=r>1$, i.e.
\begin{equation}
	\bm{X} = \sum_{j=1}^{r} \lambda_j\bm{\psi}_j\bm{\psi}_j^T,
	\label{X}
\end{equation}
where $(\lambda_j,\bm{\psi}_j)$ are the non-trivial eigenpairs of $\bm{X}$ arranged in descending order, such that $\lambda_1\geq \lambda_2 \geq ...\geq	\lambda_r>0$. In the extraordinary case that rank$(\bm{X})=1$, the relation $\bm{X} = \lambda_1\bm{\psi}_1\bm{\psi}_1^T = \bm{x}\bm{x}^T$ is of course retained, but this can only occur if the primal optimization problem is convex; i.e. in practice it will never happen for the family of optimization problems we are interested in. Therefore, the simplest idea is to approximate $\bm{x}$ via the rank-1 approximation of $\bm{X}$, i.e. assume that $\bm{x}\approx\sqrt{\lambda_1}\bm{\psi}_1$, cf. \citet{luo2010semidefinite}. An approximation of the density design field $\bm{\theta}$ can thereafter be obtained from \eqref{thetaConst2}. It is however emphasized that this $\bm{x}$ will be infeasible in our original QCQP in \eqref{optElim} in practice, and similarly for $\bm{\theta}$ in \eqref{opt2}. To obtain feasible density design variables $\bm{\theta}$, we can employ heuristics and simply map $\bm{\theta}$ to $[-1,1]$, e.g. if $\theta_j>1\mapsto\theta_j = 1$ or if $\theta_j<-1\mapsto\theta_j = -1$. 

\section{Objective functions}
\subsection{Normalized overlap}
A specific optimization problem we will address in the subsequent numerical examples is that of the maximized normalized overlap between the fields $\bm{z}$ and $\bm{c}$, $\vert\vert\bm{c}\vert\vert_2 = 1$, in the region specified by the indices given by the set $S\subseteq\{1,2,...,n\}$. In other words, we are searching for a structure that gives us a specified output response $\bm{c}$. In this case, the objective function in \eqref{obj} reduces to
\begin{equation}
	f(\bm{z}) = \frac{\bar{\bm{z}}\bm{P}\bm{z}}{\bar{\bm{z}}\bm{Q}\bm{z}},
	\label{normOverlap}
\end{equation}
where $\bm{P} = \bm{R}\bm{c}\bar{\bm{c}}\bm{R}^T$, $\bm{Q} = \bm{R}\bm{R}^T$, $\bm{p}=\bm{q} =\bm{0}$, $r=s=0$ and the matrix $\bm{R}\in\mathbb{R}^{n\times
		n}$ is diagonal with
\begin{equation}
	R_{jj} = \begin{cases}
		1, \ \ \text{if} \ \ j\in S \\[10pt]
		0, \ \ \text{otherwise}
	\end{cases} \ \ \text{for} \ \ j = 1,...,n.
	\label{R}
\end{equation}

\subsection{Normalized magnitude}
By instead equating $\bm{P} = \bm{R}\bm{c}\bar{\bm{c}}\bm{R}^T$, $\bm{Q} = \bm{0}$, $\bm{p}=\bm{q} =\bm{0}$, $r=0$ and $s=1$ the objective in \eqref{obj} becomes
\begin{equation}
	f(\bm{z}) = \bar{\bm{z}}\bm{P}\bm{z},
	\label{Power}
\end{equation}
which can be interpreted as the output overlap magnitude $\vert\bar{\bm{c}}\bm{z}\vert^2$ in region $S$. If $\bm{c} = \bm{1}$, the objective function instead simply promotes a maximized field magnitude squared in the region defined by $S$. 

\section{Numerical examples}
We solve all optimization problems using an implementation in MATLAB. As
already mentioned, the topology optimization problem in \eqref{Topopt1} is solved in a staggered fashion using MMA (\citet{svanberg1987method}). The performance bound SDP in \eqref{optRelaxSDP} is solved using Mosek (\citet{aps2019mosek}) in CVX, a package for specifying and solving convex programs (\citet{grant2008graph,grant2014cvx}). To solve the equilibrium equations, we employ a traditional finite element method. Unless stated otherwise, the topology optimization algorithm is terminated if the maximum change in a single design variable between two sequential optimization iterations is smaller than $10^{-3}$ or if the number of design iterations exceeds $1000$.

\subsection{Mode converter}
\begin{figure}[H]
	\centering
	\includegraphics[width=0.8\textwidth]{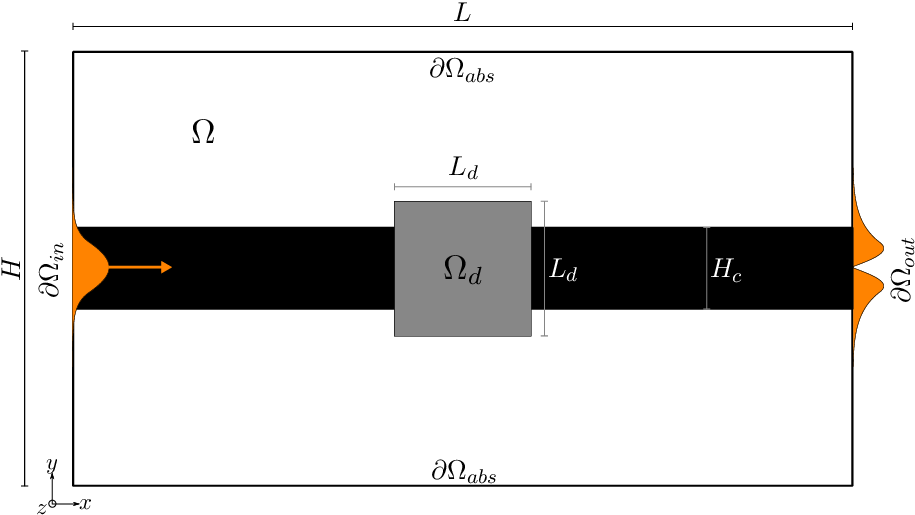}
	\caption{The mode converter geometry. The relative permittivity is fixed in white and black regions. The design domain $\Omega_d$ is the centered gray square in which the permittivity is to be determined.}
	\label{modeConverter}
\end{figure}

In the first example, we apply \eqref{opt1} to a photonic mode converter problem, similarly to \citet{angeris2023bounds}. We limit ourselves to 2D and transverse-magnetic (TM) polarisation of the electric field $\bm{E}$, i.e. $\bm{E} = [0,0,E_z]^T$. In this way, the Maxwell equations allow the derivation of the scalar-field Helmholtz equation
\begin{equation}
	\bm{\nabla}\cdot\bm{\nabla}E_z + k^2\varepsilon_r E_z = 0,
	\label{maxwellTM}
\end{equation}
for the electric field, where $\varepsilon_r$ is the relative permittivity and $k$ is the wavevector. After solving \eqref{maxwellTM}, the magnetic field may be recovered using Maxwell's equations. The mode converter geometry is depicted in Fig. \ref{modeConverter}.
\noindent
We apply first-order absorbing boundary conditions on $\partial\Omega_{abs}\subset\partial\Omega$, i.e. $\bm{\nabla}E_z\cdot \bm{n} = -ik E_z$, where $\bm{n}$ is the outwards unit normal to $\partial\Omega$. On $\partial\Omega_{in}$ and $\partial\Omega_{out}$, we excite the input mode and extract the output mode, respectively. This confines the set $S$ to the degrees-of-freedom on the right hand side of the domain, cf. \eqref{R}. The relevant eigenmodes are computed at the boundaries $\partial\Omega_{in}$ and $\partial\Omega_{out}$ using 1D line elements, assuming perfect conductors at the end-points.

The geometry in Fig. \ref{modeConverter} is discretized by four node quadrilateral finite elements. The finite element formulation of \eqref{maxwellTM} results in a linear system of equations which may be written as
\begin{equation}
	\bm{A}(\bm{\theta})\bm{z} = (\bm{K} + i\bm{D}(k) - k^2\bm{M}(\theta))\bm{z} = \bm{b},
	\label{FEmode1}
\end{equation}
where we refer to \citet{christiansen2023inverse} and the Appendix for more details on the derivations and involved matrices. It is emphasized that $\bm{M}(\theta)$ is a diagonal \say{mass matrix}, where $\bm{M}^o$ is HRZ lumped, see the Appendix for details.

In this example, the design dependency is introduced such that
\begin{equation}
	\bm{M}(\theta) = \text{diag}\left(\frac{1}{2}(\bm{\theta}^T(\varepsilon_r-1) + (\varepsilon_r+1)) \bm{M}^o\right).
	\label{mc_M}
\end{equation}
In this way, the density design variables interpolate between a high index material with $\varepsilon_r=10$ (black), and air with $\varepsilon=1$ (white). The density design variables in $\Omega_d$ are initially set to zero, whereas they are fixed to unity inside the waveguide channel (black regions) and negative unity outside the channel (white regions), cf. Fig. \ref{modeConverter}.

The parameters used in this problem are summarized in Tab. \ref{tab:mode}, unless stated otherwise.
\begin{table}[H]
	\centering
	\begin{tabular}{lr}
		\hline
		Dimensions of $\Omega$ $(L,H)$ [$\mu$m]  & $(4,2)$         \\
		Finite element mesh size                 & $336\times 168$ \\
		Free-space wavelength $\lambda$ [$\mu$m] & $0.8$           \\
		Relative permittivity $\epsilon_r$ [-]   & $10$            \\
		Dimension of $\Omega_d$ $(L_d)$ [$\mu$m] & $0.5$           \\
		Height of channel $H_c$ [$\mu$m]         & $1/6$           \\
		\hline
	\end{tabular}
	\caption{The parameters used in the mode converter example.}
	\label{tab:mode}
\end{table}
\noindent
The topology optimization only takes place in $\Omega_d\subset\Omega$; in $\Omega\setminus\Omega_d$ the relative permittivity is fixed. In this way, we can utilize a reduction of the design space when computing the bounds, cf. the \emph{Passive regions} subsection in the Appendix.

\subsubsection{Normalized mode-shape overlap}
\begin{figure}[H]
	\centering
	\includegraphics[width=0.8\textwidth]{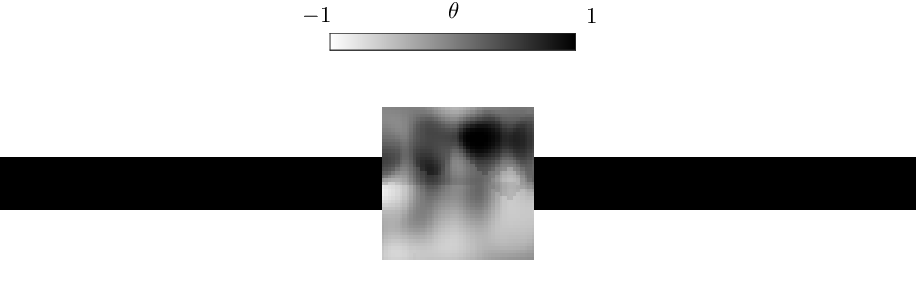}
	\caption{The topology optimized design when maximizing the normalized mode overlap.}
	\label{fig:mc_eff_designs}
\end{figure}

First, the normalized mode overlap objective in \eqref{normOverlap} is investigated. The input mode on the left hand side of the domain is the first-order mode of the waveguide, whereas the output target mode on the right hand side is the second-order mode. In Fig. \ref{fig:mc_eff_designs}, the topology optimized design is shown. Our termination criterion was reached after 170 design iterations.  

\begin{figure}[H]
	\centering
	\includegraphics[width=0.6\textwidth]{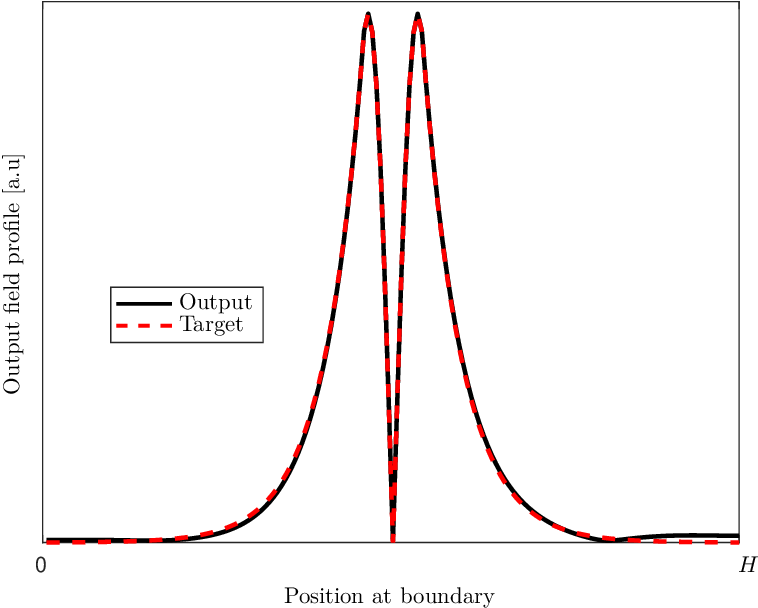}
	\caption{The normalized output $\vert E_z\vert/\vert\vert E_z\vert\vert_2$ and target mode profiles at the right-hand side boundary for the Fig. \ref{fig:mc_eff_designs} design. }
	\label{fig:mc_eff_output}
\end{figure}

\begin{figure}[H]
	\centering
	\includegraphics[width=0.7\textwidth]{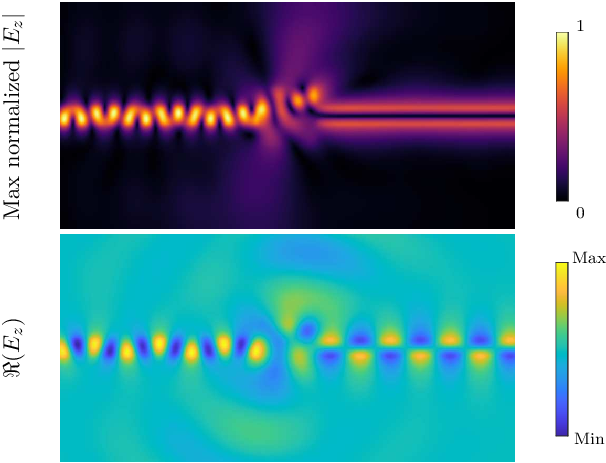}
	\caption{The (top) max normalized $\vert E_z\vert$ and (bottom) real part $\Re(E_z)$ over $\Omega$ corresponding to the Fig. \ref{fig:mc_eff_designs} design. }
	\label{fig:mc_eff_fields}
\end{figure}

The mode purity, i.e. objective function, corresponding to the Fig. \ref{fig:mc_eff_designs} design is $f = 0.9993$ for the topology optimization problem in \eqref{Topopt1} and $1.0000$ for the bound problem in \eqref{optRelaxSDP}; the relative difference is $0.07\%$. Hence, the bounds provide the a prior unknown knowledge that perfect mode-conversion is indeed possible given the chosen design freedom and further we find that topology optimization identifies the optimal solution to within what is by all practical purposes an acceptable margin of error starting from a single initial guess. We plot the output and target mode profiles at the right-hand side boundary in Fig. \ref{fig:mc_eff_output}.
In Fig. \ref{fig:mc_eff_fields}, the $E_z$ field is plotted over $\Omega$. Since the objective function being maximized is the mode-shape overlap, thus neglecting the output power, we emphasize that the observable reduction in field amplitude as well as the reflection back into the input waveguide and scattering is not unexpected. To further justify this claim, we compute the power transmittance $T$ in the second-order mode via the time-averaged Poynting vector, cf. \citet{christiansen2023inverse}. The Fig. \ref{fig:mc_eff_designs} design has a transmittance of only $T \approx 58\%$.

\subsubsection{Mode overlap magnitude}
\begin{figure}[H]
	\centering
	\includegraphics[width=0.9\textwidth]{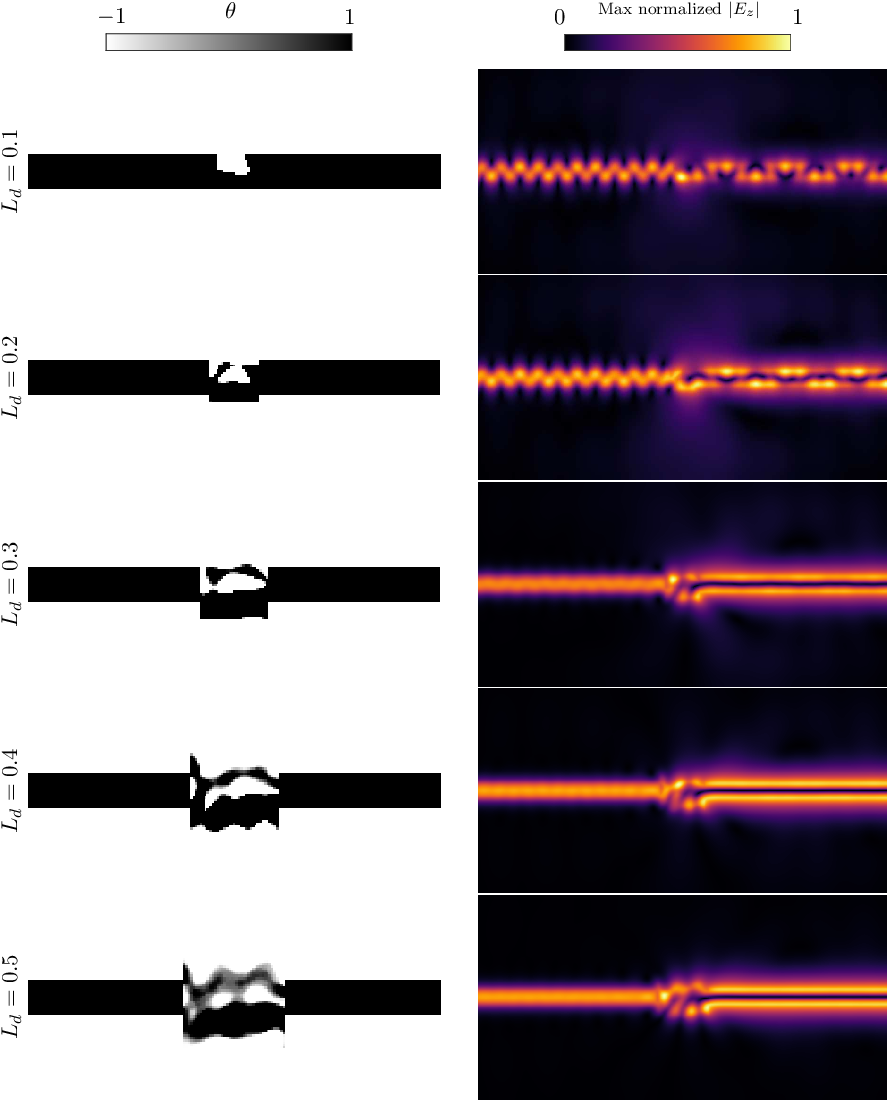}
	\caption{The topology optimized designs (left) and row corresponding max-normalized field distributions (right) for varying design domain sizes when maximizing the mode overlap magnitude. With reference to the designs starting from the top, the optimization terminated at 235, 684, 1000, 1000 and 1000 design iterations, respectively.}
	\label{fig:mc_mag_designs_diffD}
\end{figure}

To obtain a mode converter with higher transmittance, i.e. one that produces the target mode output but also maximizes the output magnitude, we next choose the maximum mode overlap magnitude objective in \eqref{Power}. Again, the input mode on the left hand side is the first-order mode, whereas the output target on the right hand side is the second-order mode including its amplitude, i.e. $m=2$, cf. \eqref{Power}. In this example, we vary the size of the design domain $L_d\in\{0.1667, 0.2381, 0.3333, 0.4286, 0.5000\}$ $\mu$m to investigate how this influences the design and bounds. The topology optimized designs and corresponding field distributions are depicted in Fig. \ref{fig:mc_mag_designs_diffD}.

\begin{figure}[H]
	\centering
	\includegraphics[width=\textwidth]{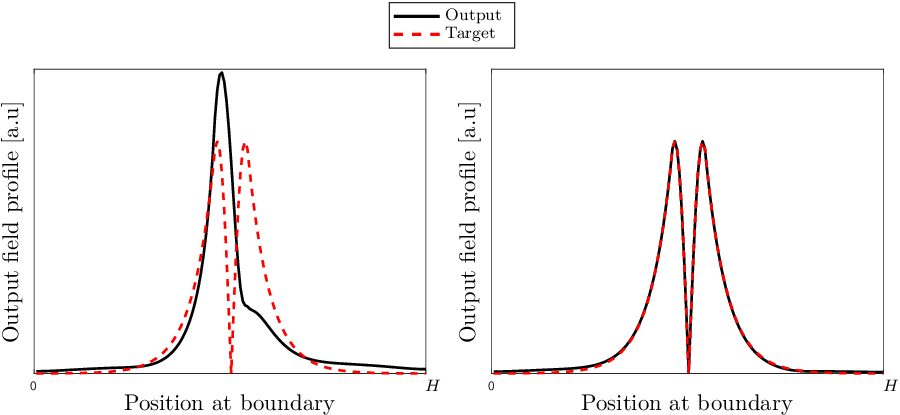}
	\caption{The normalized output $\vert E_z\vert/\vert\vert E_z\vert\vert_2$ and target mode profiles at the right-hand side boundary for the Fig. \ref{fig:mc_mag_designs_diffD} top and bottom designs. }
	\label{fig:mc_mag_output}
\end{figure}

Since the objective is posed such that we still aim to match the target output mode profile, we plot the normalized output and target mode profiles at the right-hand side boundary for the smallest and largest $L_d$ designs in Fig. \ref{fig:mc_mag_output}. The larger design domain clearly matches the target waveform profile better than the smaller design domain, as we expected based on our basic understanding of the physical system and previous numerical results.

\begin{figure}[H]
	\centering
	\includegraphics[width=\textwidth]{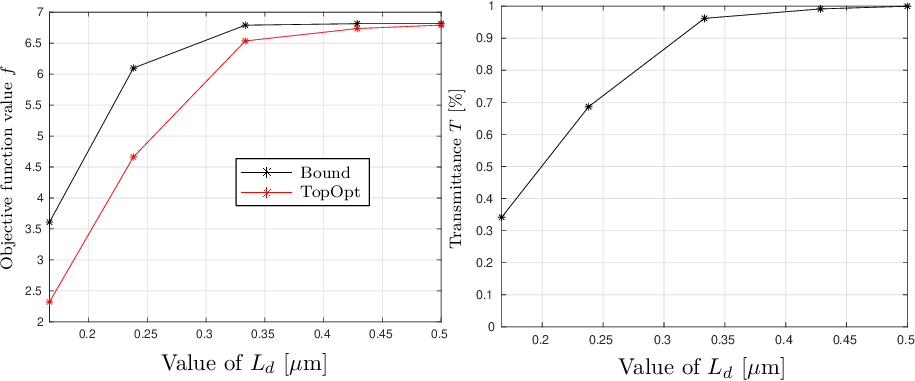}
	\caption{Topology optimization objective function and bound (left) or transmittance (right) versus the domain dimension corresponding to the Fig. \ref{fig:mc_mag_designs_diffD} designs.}
	\label{fig:mc_mag_plot_diffD}
\end{figure}

\begin{figure}[H]
	\centering
	\includegraphics[width=0.6\textwidth]{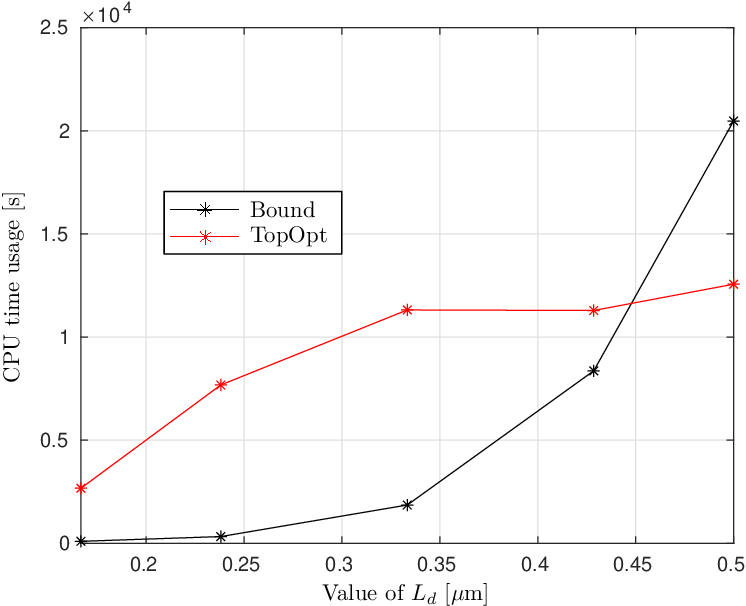}
	\caption{The time usage in seconds for the topology optimization and bound problems. }
	\label{fig:mc_time}
\end{figure}

In Fig. \ref{fig:mc_mag_plot_diffD} we plot both the topology optimization and bound objective values and transmittance versus the range of $L_d$. As expected, the objective function plateaus when the design domain is sufficiently large since the transmittance approaches unity. We also emphasize that for the largest design domain, the topology optimization is able to design a \say{perfect} mode converter starting from a single initial guess; the gap between the bound and topology optimization objective values is as small as $0.33\%$ for the largest design domain. For the smaller design domains, it is noted that the topology optimization does not generate perfect mode converters, but by comparing to the bound we realize that this should not be possible either. This illustrates the power of the bounds, both as a predictive tool a priori telling the designer what is possible, and as a tool explaining why a particular design is not performing better despite being optimized. We further tried to recompute the bounds without the cross-correlation constraints in \eqref{thetaConst1}. However, this always resulted in less tight bounds; even unbounded results could be obtained, illustrating the importance of including \eqref{thetaConst1} in the bounds formulation.
In Fig. \ref{fig:mc_time}, we plot the CPU time usage for the topology and bound optimization problems versus the design region size $L_d$, which directly translates to the design variable space size. The topology optimization problems for $L_d \in\{0.1667,0.2381\}$ $\mu$m terminated earlier than the rest; at 235 and 684 design iterations respectively. The remaining topology optimization runs terminated at 1000 design iterations and have similar time usage. This is expected since the number of degrees-of-freedom is constant for all choices of $L_d$; it is not the MMA optimizer that poses the greatest bottleneck in the topology optimization problem; it is the solution of the state problem. Fig. \ref{fig:mc_time} illustrates a key challenge with the bound problem; its time usage scales poorly with the number of design variables. Already at $L_d = 0.5$ $\mu$m, which corresponds to 1849 design variables, the time to solve the bound optimization problem passes the time to solve the topology optimization problem. The graph clearly illustrates the exponential time usage of the bound problem. We estimated the time dependency to scale with $\thicksim n_{\theta_f}^3$, where $\theta_f$ is the number of designable degrees-of-freedom.

\begin{figure}[H]
	\centering
	\includegraphics[width=0.8\textwidth]{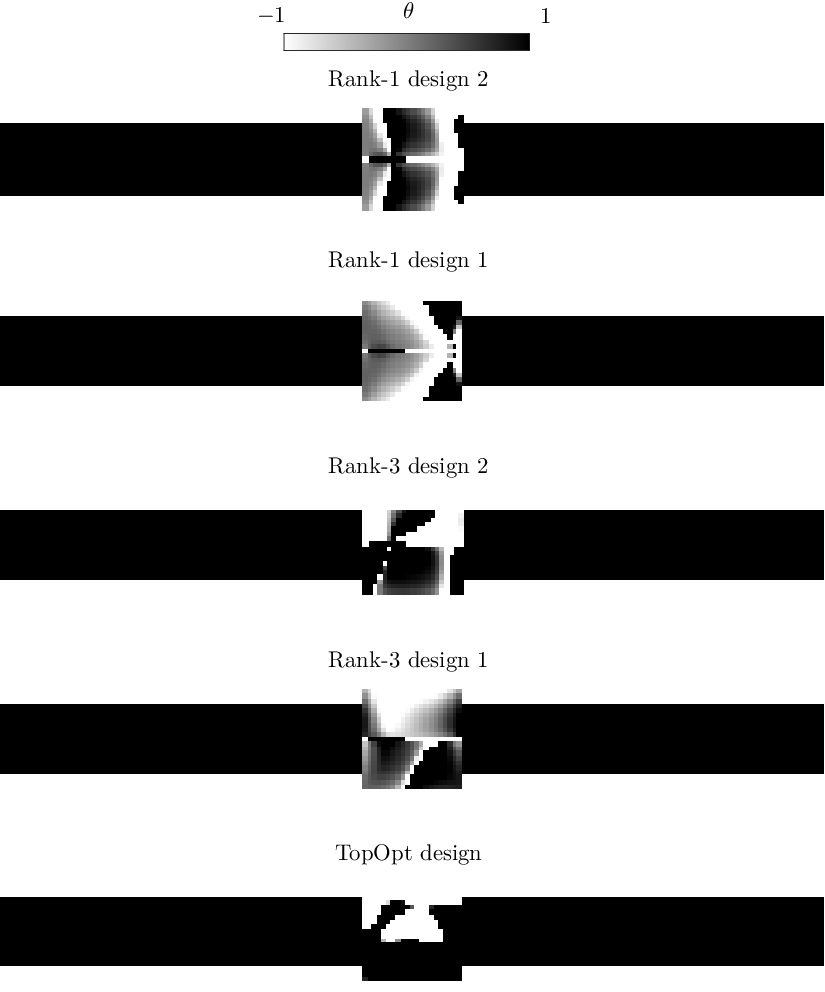}
	\caption{The rank approximations of the density design obtained from the bound optimization problem.}
	\label{fig:mc_bound_design}
\end{figure}

The bound optimization problem for $L_d = 0.2381$ $\mu$m yields a solution with rank$(\bm{X})\approx 3$. In Fig. \ref{fig:mc_bound_design} we plot the rank-1 and rank-3 approximations of the density design field, i.e. when we approximate $\bm{x}\approx\sum_{j=1}^r \sqrt{\lambda_j}\bm{\psi}_j$ using $r=1$ or $r=3$, respectively. We emphasize that the cross-correlation constraints in \eqref{thetaConst1} enforce $\theta_{n+j}'=\theta_j'$, $j=1,2,...,n$, i.e. the first and second half's of $\bm{\theta}'\in\mathbb{R}^{2n}$ must equal. However, as long as rank$(\bm{X})\neq1$, the rank approximations must not fulfill these constraints, since the constraint is explicitly enforced on the \emph{matrix} $\bm{X}$ via \eqref{thetaConst5}, not on the sum of the vectors $\bm{\psi}_j$, $j=1,...,r$, from which the approximations of $\bm{\theta}'$ can be computed, cf. \eqref{X}. Therefore, Fig. \ref{fig:mc_bound_design} depicts two designs for each rank approximation corresponding to the first and second half's of $\bm{\theta}'$. Clearly, the two rank-1 approximative designs have no resemblance to the topology optimized design at all, which exposes the non-convexity of the primal optimization problem. These designs are dysfunctional, with post-processing objective function values of $\approx 0$. Indeed, this is not very surprising since the structures in Fig. \ref{fig:mc_bound_design} are symmetric with respect to the middle $x$-plane, which clearly suppresses the ability to transform the symmetric first-order excitation mode to the asymmetric second-order target mode (Fig. \ref{fig:mc_eff_fields} bottom). The two rank-3 approximative designs are not symmetric with respect to the middle $x$-plane, wherefore their post-processing objective function values are greatly improved compared to the rank-1 approximative designs. We also tried to use these designs as initial guesses for the topology optimization to see whether it could improve the design, but the same local minima as depicted in Fig. \ref{fig:mc_mag_designs_diffD} was found. We believe this can be caused by the relatively low rank of $\bm{X}$.

\begin{figure}[H]
	\centering
	\includegraphics[width=0.6\textwidth]{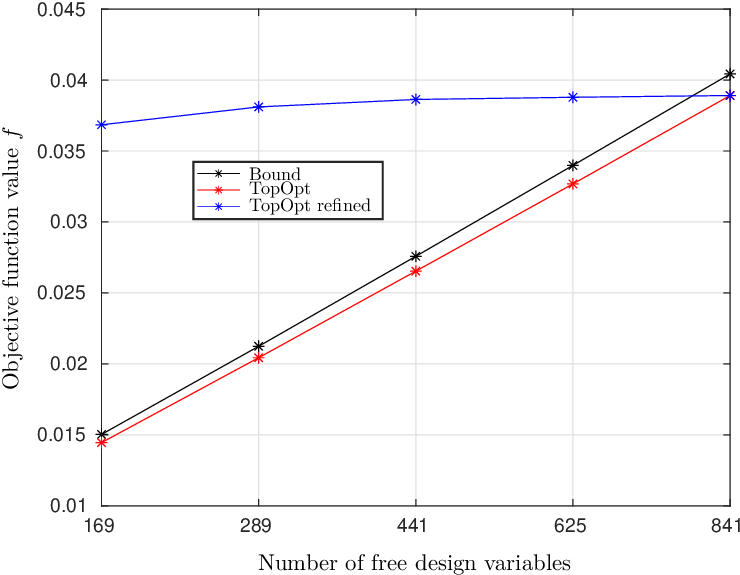}
	\caption{The objective function of the topology (red) and bound (black) optimization problems versus the discretization represented by the number of free design variables. The blue graph represents the objective function value of the topology optimization design when interpolated on the finest discretization.}
	\label{fig:mc_meshstudy}
\end{figure}

The solution to the bound optimization problem and thus the bound value will depend on the discretization of the physics. To illustrate this, we plot the objective function of the topology and bound optimization problems versus the number of free design variables when using $L_d=0.3333$ $\mu$m in Fig. \ref{fig:mc_meshstudy}. In this case, the objective function values are normalized by the number of degrees-of-freedom in the $y$-direction. In the same figure, we also plot the topology optimization objective function calculated for the increasingly refined designs evaluated on the finest mesh. While the bound (black curve) is respected for each topology optimized design when evaluated on the meshes that they where optimized on (red curve), it is seen that the coarse topology optimized designs  evaluated on the fine mesh (blue curve) actually surpass the predicted bound, which at a first glance should not be possible. This is however not surprising, yet a key point, when considering that the bounds have been computed for a particular problem size (mesh resolution), and thus are only guaranteed to be valid for said mesh resolutions. This example illustrates that one must exercise great care when computing bounds for a particular problem; the bounds are mathematical/computational in nature and thus relates to the particular optimization problem being solved and only make physical sense if the physics is sufficiently resolved. 

Although the design in the bottom panel of Fig. \ref{fig:mc_mag_designs_diffD} exhibits almost perfect mode conversion and transmittance ($T\approx 0.9996$), it consists of large intermediate density regions wherein $\theta\in(-1,1)$. This hinders the physical realization of the design as it is challenging if not impossible to realize intermediate $\varepsilon_r$ values at such small length scales. Therefore, to end this section, we implement a \say{conventional} topology optimization approach with penalization, filtration and thresholding, such that intermediate density regions are penalized. The filtering is a traditional convolution filter (\citet{bruns2001topology}, \citet{bourdin2001filters})
\begin{equation}
	\tilde{\theta}_j = \frac{\ds\sum_{k\in N_j} H_{jk}\theta_j}{\ds\sum_{k\in N_j} H_{jk}},
	\label{filter}
\end{equation}
with a cone shaped kernel such that $H_{jk} = \text{max}(0,r-\Delta_{jk})$, where $N_j$ is the set of nodes $k$ for which the node-to-node distance $\Delta_{jk}$ to node $j$ is smaller than $r$. The thresholding is performed via an approximate Heaviside projection around zero (\citet{wang2014interpolation})
\begin{equation}
	\bar{\theta}_j = H_{\beta}(\tilde{\theta}_j) =  \frac{\text{tanh}(\beta\tilde{\theta}_j)}{\text{tanh}(\beta)}.
	\label{Heavi}
\end{equation}
The penalization follows a so-called pamping scheme (\citet{jensen2005topology}), wherein intermediate density regions are penalized by introducing a density dependent attenuation, i.e.
\begin{equation}
	\varepsilon_r \leftarrow \frac{1}{2}(\bar{\theta}_j(\varepsilon_r-1) + (\varepsilon_r+1)) - i\eta(\bar{\theta}_j+1)(1-\bar{\theta}_j),
	\label{pamping}
\end{equation}
where $\eta=1$ is a scaling parameter. In this way, intermediate densities $\theta\in(-1,1)$ are penalized since the objective is to maximize the output field magnitude, i.e. the optimizer will tend to avoid attenuating materials.

\begin{figure}[H]
	\centering
	\includegraphics[width=\textwidth]{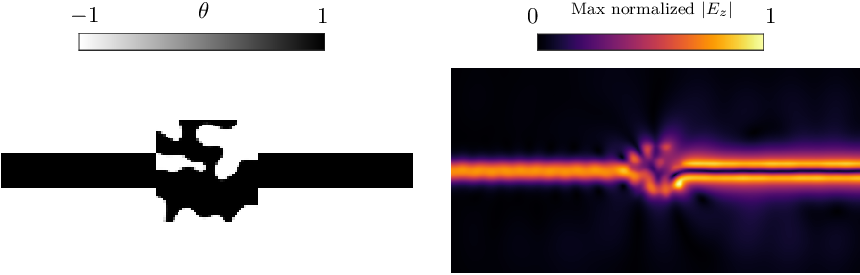}
	\caption{The gray-scaled penalized topology optimization design (left) and corresponding max normalized field distribution (right).}
	\label{fig:mc_topopt}
\end{figure}

We use a filter radius $r$ of three times the element length. The Heaviside projection slope parameter $\beta$ is initially $\beta=1$ and increased $\beta \leftarrow 2\beta$ every 50th iteration until $\beta = 8$. The topology optimized design obtained using the conventional framework is depicted in Fig. \ref{fig:mc_topopt}. In a post-processing evaluation without pamping ($\eta = 0$ in \eqref{pamping}), the Fig. \ref{fig:mc_topopt} design has a transmittance of $T\approx 0.9685$, i.e. by obtaining a realizable design we only drop approximately $3\%$ in performance. We note that this performance drop can be reduced significantly by allowing added design freedom using a finer finite-element discretization (\citet{christiansen2023inverse}) and possibly continuation in $\eta$.

\subsection{Resonating elastic plate}
\begin{figure}[H]
	\centering
	\includegraphics[width=0.6\textwidth]{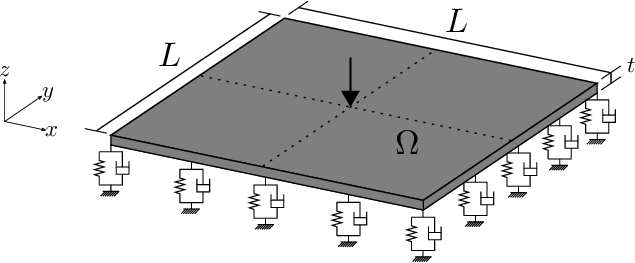}
	\caption{The 2D plate geometry.}
	\label{fig:plate}
\end{figure}

In the next example, we apply \eqref{opt1} to a resonating
elastic plate problem depicted in Fig. \ref{fig:plate}. This problem is inspired by the quantum optomechanical resonator example in \citet{gao2020systematic}.
\noindent

The plate is discretized using 2D Kirchhoff plate elements. Damping boundary conditions are applied on all boundaries of the plate, which are modeled using a parallel assembly of a dashpot and an elastic spring, such that the imaginary vertical force
\begin{equation}
	F_b = s(1+i\gamma)w,
	\label{damping_bc}
\end{equation}
is applied to all boundaries. In the above, $s$ is the spring stiffness, $s\gamma$ controls the damping and $w$ is the out-of-plane displacement.

Similarly to the previous example, the equilibrium equations may be written on the form
\begin{equation}
	\bm{A}(\bm{\theta})\bm{z} = (\bm{K} + \bm{C} + i\bm{D} - \omega^2\bm{M}(\theta))\bm{z} = \bm{b},
	\label{plateFEmode1}
\end{equation}
where $\bm{K}\in\mathbb{R}^{n\times n}$ is the symmetric elastic stiffness matrix, $\bm{M}\in\mathbb{R}^{n\times n}$ is the diagonal lumped mass matrix and $\omega$ is the excitation frequency. Finally, the $\bm{C}$ and $\bm{D}$ matrices contain the contributions from the boundary conditions in \eqref{damping_bc}. It is emphasized that the use of Kirchhoff plate elements entails the existence of three degrees-of-freedom in each node; one out-of-plane displacement degree-of-freedom and two rotational degrees-of-freedom, making it a vector-field problem.

In this example, the design dependency is introduced such that
\begin{equation}
	\bm{M}(\theta) = \frac{1}{2}\left(\bm{\theta}(\bar{\rho}-\underline{\rho}) + (\bar{\rho}+\underline{\rho})\right)\bm{M}^o,
	\label{massScaling}
\end{equation}
where $\bar{\rho}$ and $\underline{\rho}$ are the upper and lower bound mass densities and $\bm{M}^o$ is the HRZ lumped mass matrix, cf. the Appendix. Physically, this can be interpreted as added point masses (dimples) that do not influence the plate stiffness. The density design variables in $\Omega$ are initially set to zero. Since $\bm{\theta}\in \mathbb{R}^{n}$, we prescribe all $\theta_j$ associated with the rotational degrees-of-freedom $j=2,3,5,6,...,n-1,n$ to zero, i.e. we again utilize the reduction of the design space that is described in the Appendix. It is emphasized that this prescribed value of $\theta_j$ can be set arbitrarily since the HRZ lumping procedure involves setting all diagonal elements of $\bm{M}^o$ pertaining to the rotational degrees-of-freedom to zero, cf. the Appendix.

The parameters used in this problem are summarized in Tab.
\ref{tab:plate}.
\begin{table}[H]
	\centering
	\begin{tabular}{lr}
		\hline
		Dimensions of $\Omega$ $(L,t)$ [m] & $(0.5,0.01)$ \\
		Young's Modulus $E$ [kPa]                                 & $1$         \\
		Poisson's ratio $\nu$ [-]                                 & $0.4$       \\
		Mass densities $(\underline{\rho},\bar{\rho})$ [kg/m$^3$] & $(10,10^2)$ \\
		Spring stiffness $k$ [N/m$^2$]                            & $10^{-2}$   \\
		Damping parameter $\gamma$ [-]                            & $50$        \\
		\hline
	\end{tabular}
	\caption{The parameters used in the resonating plate example.}
	\label{tab:plate}
\end{table}

We again consider the maximum amplitude magnitude objective, cf. \eqref{Power}. The amplitude should be maximized in the center out-of-plane degree-of-freedom, in which the harmonic loading is applied. We now keep the discretization fixed at $30\times 30$ elements, and vary the excitation frequency and investigate how the topology optimization results compares to the bound. The relatively coarse mesh is needed due to the extensive memory requirements of the bound problem.

\begin{figure}[H]
	\centering
	\includegraphics[width=\textwidth]{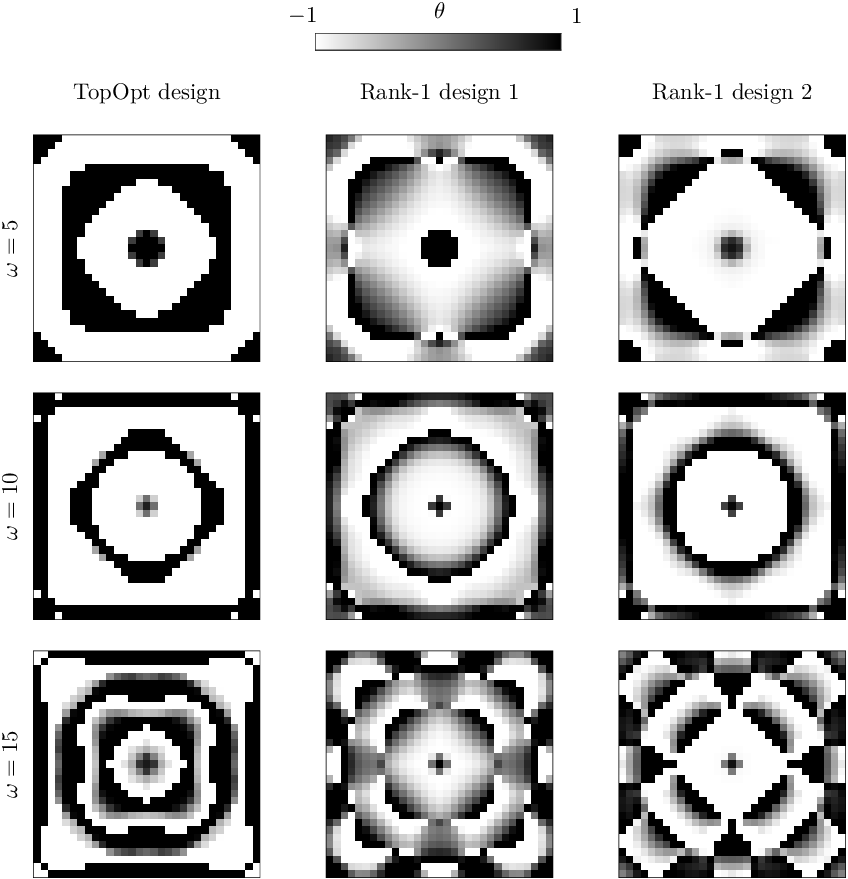}
	\caption{The topology optimized and rank-1 approximated designs for varying excitation frequencies. With reference to the designs starting from the top, the optimization terminated at 179, 1000, 1000 design iterations, respectively.}
	\label{fig:plate_designs}
\end{figure}

\begin{figure}[H]
	\centering
	\includegraphics[width=\textwidth]{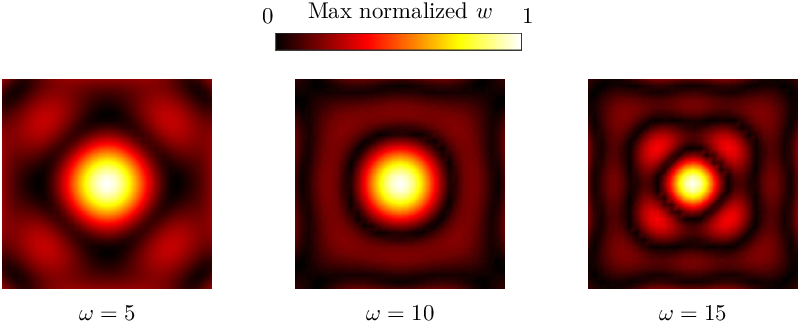}
	\caption{The max normalized out-of-plane displacement fields $w$ corresponding to the topology optimized designs in the first column of Fig. \ref{fig:plate_designs}.}
	\label{fig:plate_fields}
\end{figure}

\begin{figure}[H]
	\centering
	\includegraphics[width=\textwidth]{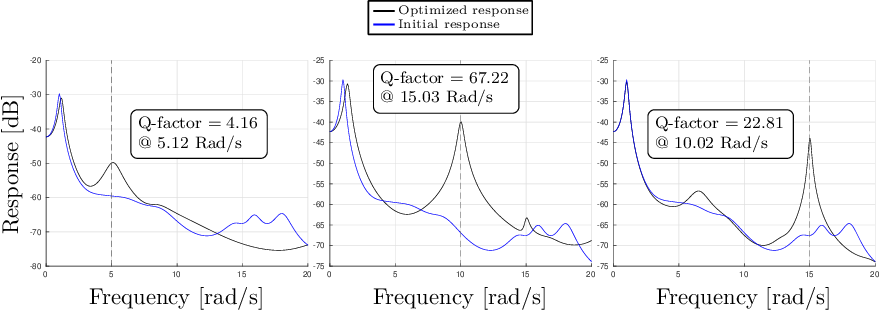}
	\caption{The initial and optimized frequency response of the Fig. \ref{fig:plate_designs} topology optimized designs. The vertical dashed black lines highlights the excitation frequencies.} 
	\label{fig:plate_FR}
\end{figure}

In Fig. \ref{fig:plate_designs}, the topology optimized designs of the resonating plate are depicted for varying excitation frequencies. In the same figure, we also include plots of the rank-1 approximated designs obtained from the bound problems. Interestingly, we identify certain similarities between the topology optimization and rank-1 approximated bound designs. For all presented examples in Fig. \ref{fig:plate_designs} the rank$(\bm{X})\gg 1$.
The max normalized out-of-plane field $w$ pertaining to the Fig. \ref{fig:plate_designs} topology optimized designs are shown in Fig. \ref{fig:plate_fields}. It is confirmed that the designs have the largest out-of-plane field amplitudes at the center of the plates.
In Fig. \ref{fig:plate_FR}, the frequency response of the initial design is compared to that of the topology optimized design. The response is calculated as the average magnitude of the out-of-plane displacement squared. It is confirmed that the optimized designs have increased the output response at the target excitation frequency compared to the initial response. In the same figure, we also include the Q-factors of the resonators, cf. \citet{gao2020systematic} for more details on the computations.
In Tab. \ref{tab:plate_results}, the objective function values of the topology optimization are compared to the bound. 
\begin{table}[H]
	\centering
	\begin{tabular}{c|c|c|c}
		Frequency [rad/s] & TopOpt $[\cdot 10^{-4}]$ & Bound $[\cdot 10^{-4}]$ & Relative gap size [$\%$] \\
		\hline
		$5$               & $3.18$                   & $6.85$                  & $53.64$                  \\
		$10$              & $56.10$                  & $128.72$                & $56.42$                  \\
		$15$              & $36.81$                  & $679.04$                & $94.58$                  \\
	\end{tabular}
	\caption{The topology optimization and bound objective values for different excitation frequencies. }
	\label{tab:plate_results}
\end{table}
\noindent
It is seen that the $\omega=5$ rad/s and $\omega=10$ rad/s designs are closer to their bound than the $\omega=15$ rad/s design. Nonetheless, they are still quite far from the bound. Using the rank-1 approximations of the $\omega=5$ rad/s and $\omega=10$ rad/s designs as initial guess for the optimization does not get us closer to the bound, which is not very surprising since the rank-1 designs look similar to the topology optimized design, cf. Fig. \ref{fig:plate_designs}. We also tried using higher rank approximations as initial guesses but could not get closer to the bound.

For the $\omega=15$ rad/s problem, we first tried to use a range of uniform design guesses for the topology optimization to see whether a better optima could be located. This did however only lead to a slightly improved design. Next, we tried using rank-1 to rank-7 approximated designs as initial guesses. The best design for $\omega=15$ rad/s was obtained when using the rank-4 approximated design 2 as initial guess for the topology optimization and changing the termination criteria to 10000 design iterations (slow convergence was observed), but this only reduced the relative gap size to $\approx 88\%$. In these examples, since rank$(\bm{X})\gg 1$ we can question the tightness of the bounds, as these are computed by disregarding the rank$(\bm{X})=1$ constraint.

\begin{figure}[H]
	\centering
	\includegraphics[width=0.6\textwidth]{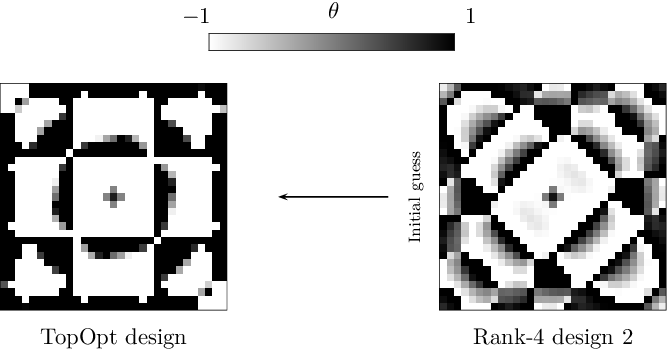}
	\caption{The topology optimized design (left) obtained when using the rank-4 approximated design 2 (right) as initial guess.} 
	\label{fig:plate_15}
\end{figure}

While the topology optimization tool is capable of solving this problem on significantly finer meshes, unfortunately the CPU time and memory demands currently hinders such studies when computing the bounds. In the future, more efficient algorithms (e.g. \citet{kim2011exploiting} or \citet{yurtsever2021scalable}) should be implemented when solving larger-scale bound problems; hopefully this would drastically increase the methods efficacy.

\section{Conclusions}
In this work, we have presented a framework for computing performance bounds to a class of topology optimization problems. We have developed and extended the framework by \citet{angeris2023bounds} which relies on the finite difference method, to incorporate cross-correlation constraints for stability and finite element discretization methods for generality. The computational framework is tested on two topology optimization problems; 1) a photonic mode converter and 2) an elastic resonating plate. To the author's knowledge, the latter problem pioneers the investigation of computational performance bounds for both vector-field and linear elastic inverse design problems.

The numerical examples illustrate the power of computational performance bounds in topology optimization. By having access to the bounds, we know when the topology optimization produces \say{optimal} designs, and when we should rerun the optimization with another initial guess for the design variables. We can also verify that our topology optimization framework is able to generate near optimal designs for certain problems. However, the numerical examples also illustrate the current limitations of the computational bounds. Using the proposed solution method, we are limited to small problem sizes and certain optimization problems.

\subsection{Outlook}
Although this work has exemplified the usage of computational bounds in topology optimization, several open questions remain regarding their generality and efficiency. To end this paper, we therefore pose a handful of these questions and provide suggestions on how to go about (hopefully) to answer them:
\begin{itemize}

	\item \emph{What if the design dependency is not present in the \say{mass} matrix, but rather in the \say{stiffness} matrix, i.e. the differential operator? In other words: Is it possible to bound static problems using a modified version of the presented framework?}

	In such cases, the locality of the density design variables $\bm{\theta}$ is lost, i.e. a density design variable does in general not influence solely a single degree-of-freedom, wherefore the proposed framework for computing the bounds is not directly applicable. In such cases, 
the second-order differential equation could be converted into two coupled first order differential equations, such that design locality is preserved. For elasticity problems, we believe that this would require a mixed finite element formulation, wherein both the displacements and the stresses are treated as unknown variables. 
The investigation of such an approach is clearly an interesting subject for future work.

	\item \emph{Can arbitrary constraints possibly be included in the optimization problems and the bound?}

	The presented framework requires that the constraint is (or can be rewritten as) a quadratic (or linear) function in the state field, i.e. $\bm{z}$.

	\item \emph{Can you guarantee that the bounds are tight, i.e. define the best possible optimization outcome?}

	As also discussed by \citet{angeris2023bounds}, we cannot guarantee the tightness of the computational bounds presented herein due to the numerics involved. Numerical inconsistencies can occur due to e.g. the scaling of the problem. For example, we found it harder (in addition to more time-consuming) to solve the bound problem when considering elasticity in comparison to photonics. We attribute this issue to the large contrasts (condition numbers) that exists for the stiffness and mass matrices in the former case. In other words, the influence of the design variables is smaller in the elasticity problem. Our experience is that the computations of the bounds then become more sensitive to what we believe to be e.g. scaling and round-off errors.

	\item \emph{How can the computation of the bounds be made more efficient in relation to both CPU time and memory usage?}

	As also discussed by \citet{gertler2023many}, there is currently a growing interest in investigating the speed-up of SDP:s. For example, sparsity properties characterized by chordal graph structures can be exploited to speed-up the solution times; there already exists open-source software that utilizes such properties, cf. sparseCoLO (\citet{kim2011exploiting}). More recent work (\citet{yurtsever2021scalable}) shows promising results for solving the \say{max-cut} problem with very large SDP:s (number of entries in $\bm{X}$ up to $10^{13}$). In future work, it would be very interesting to implement and test this algorithm.

	\item \emph{Are computational bounds worth to pursue in topology optimization or is the formulation too limited for (most) practical problems?}

	We argue that the state-of-the-art computational bound frameworks are still too rudimentary for practical usage for state-of-the-art topology optimization problems. Exceptions to this statement could however be \say{simple} or small dynamical problems (such as the numerical examples included in this work). Nonetheless, if the existing frameworks are extended to be applicable to static, non-local problems, and their efficiency is increased to allow for larger problem sizes, we believe the computational bounds could be immensely useful for the topology optimization community.

\end{itemize}

\section{Acknowledgments}
This project is supported by the Villum Foundation via the Villum Investigator
project \say{AMSTRAD} (VIL54487). The authors would like to thank Dr. Owen
Miller for valuable discussions during the preparation of this work.

\section{Conflict of interest}
The authors declare that they have no conflict of interest.

\section{Replication of results} 
The authors state that all the data necessary to replicate the results are presented in the manuscript. Relevant parts of the code can be shared by contacting the corresponding author.

\bibliographystyle{elsarticle-harv}
\bibliography{database}

\begin{thebibliography}{54}
\expandafter\ifx\csname natexlab\endcsname\relax\def\natexlab#1{#1}\fi
\providecommand{\url}[1]{\texttt{#1}}
\providecommand{\href}[2]{#2}
\providecommand{\path}[1]{#1}
\providecommand{\DOIprefix}{doi:}
\providecommand{\ArXivprefix}{arXiv:}
\providecommand{\URLprefix}{URL: }
\providecommand{\Pubmedprefix}{pmid:}
\providecommand{\doi}[1]{\href{http://dx.doi.org/#1}{\path{#1}}}
\providecommand{\Pubmed}[1]{\href{pmid:#1}{\path{#1}}}
\providecommand{\bibinfo}[2]{#2}
\ifx\xfnm\relax \def\xfnm[#1]{\unskip,\space#1}\fi
\bibitem[{Aage et~al.(2017)Aage, Andreassen, Lazarov and
  Sigmund}]{aage2017giga}
\bibinfo{author}{Aage, N.}, \bibinfo{author}{Andreassen, E.},
  \bibinfo{author}{Lazarov, B.S.}, \bibinfo{author}{Sigmund, O.},
  \bibinfo{year}{2017}.
\newblock \bibinfo{title}{Giga-voxel computational morphogenesis for structural
  design}.
\newblock \bibinfo{journal}{Nature} \bibinfo{volume}{550},
  \bibinfo{pages}{84--86}.
\bibitem[{Alexandersen et~al.(2016)Alexandersen, Sigmund and
  Aage}]{alexandersen2016large}
\bibinfo{author}{Alexandersen, J.}, \bibinfo{author}{Sigmund, O.},
  \bibinfo{author}{Aage, N.}, \bibinfo{year}{2016}.
\newblock \bibinfo{title}{Large scale three-dimensional topology optimisation
  of heat sinks cooled by natural convection}.
\newblock \bibinfo{journal}{International Journal of Heat and Mass Transfer}
  \bibinfo{volume}{100}, \bibinfo{pages}{876--891}.
\bibitem[{Andreassen et~al.(2014)Andreassen, Lazarov and
  Sigmund}]{andreassen2014design}
\bibinfo{author}{Andreassen, E.}, \bibinfo{author}{Lazarov, B.S.},
  \bibinfo{author}{Sigmund, O.}, \bibinfo{year}{2014}.
\newblock \bibinfo{title}{Design of manufacturable 3d extremal elastic
  microstructure}.
\newblock \bibinfo{journal}{Mechanics of Materials} \bibinfo{volume}{69},
  \bibinfo{pages}{1--10}.
\bibitem[{Angeris(2022)}]{angeris2022note}
\bibinfo{author}{Angeris, G.}, \bibinfo{year}{2022}.
\newblock \bibinfo{title}{A note on generalizing power bounds for physical
  design}.
\newblock \bibinfo{journal}{arXiv preprint arXiv:2208.04411} .
\bibitem[{Angeris et~al.(2023)Angeris, Diamandis, Vu{\v{c}}kovi{\'c} and
  Boyd}]{angeris2023bounds}
\bibinfo{author}{Angeris, G.}, \bibinfo{author}{Diamandis, T.},
  \bibinfo{author}{Vu{\v{c}}kovi{\'c}, J.}, \bibinfo{author}{Boyd, S.P.},
  \bibinfo{year}{2023}.
\newblock \bibinfo{title}{Bounds on efficiency metrics in photonics}.
\newblock \bibinfo{journal}{ACS Photonics} \bibinfo{volume}{10},
  \bibinfo{pages}{2521--2529}.
\bibitem[{Angeris et~al.(2021)Angeris, Vu{\v{c}}kovi{\'c} and
  Boyd}]{angeris2021heuristic}
\bibinfo{author}{Angeris, G.}, \bibinfo{author}{Vu{\v{c}}kovi{\'c}, J.},
  \bibinfo{author}{Boyd, S.}, \bibinfo{year}{2021}.
\newblock \bibinfo{title}{Heuristic methods and performance bounds for photonic
  design}.
\newblock \bibinfo{journal}{Optics Express} \bibinfo{volume}{29},
  \bibinfo{pages}{2827--2854}.
\bibitem[{Angeris et~al.(2019)Angeris, Vu{\v{c}}kovi{\'c} and
  Boyd}]{angeris2019computational}
\bibinfo{author}{Angeris, G.}, \bibinfo{author}{Vu{\v{c}}kovi{\'c}, J.},
  \bibinfo{author}{Boyd, S.P.}, \bibinfo{year}{2019}.
\newblock \bibinfo{title}{Computational bounds for photonic design}.
\newblock \bibinfo{journal}{ACS Photonics} \bibinfo{volume}{6},
  \bibinfo{pages}{1232--1239}.
\bibitem[{Bathe(2006)}]{bathe2006finite}
\bibinfo{author}{Bathe, K.J.}, \bibinfo{year}{2006}.
\newblock \bibinfo{title}{Finite element procedures}.
\newblock \bibinfo{publisher}{Klaus-Jurgen Bathe}.
\bibitem[{Bends{\o}e(1989)}]{bendsoe1989optimal}
\bibinfo{author}{Bends{\o}e, M.P.}, \bibinfo{year}{1989}.
\newblock \bibinfo{title}{Optimal shape design as a material distribution
  problem}.
\newblock \bibinfo{journal}{Structural optimization} \bibinfo{volume}{1},
  \bibinfo{pages}{193--202}.
\bibitem[{Bends{\o}e and Kikuchi(1988)}]{bendsoe1988generating}
\bibinfo{author}{Bends{\o}e, M.P.}, \bibinfo{author}{Kikuchi, N.},
  \bibinfo{year}{1988}.
\newblock \bibinfo{title}{Generating optimal topologies in structural design
  using a homogenization method}.
\newblock \bibinfo{journal}{Computer methods in applied mechanics and
  engineering} \bibinfo{volume}{71}, \bibinfo{pages}{197--224}.
\bibitem[{Bluhm et~al.(2021)Bluhm, Sigmund and Poulios}]{bluhm2021internal}
\bibinfo{author}{Bluhm, G.L.}, \bibinfo{author}{Sigmund, O.},
  \bibinfo{author}{Poulios, K.}, \bibinfo{year}{2021}.
\newblock \bibinfo{title}{Internal contact modeling for finite strain topology
  optimization}.
\newblock \bibinfo{journal}{Computational Mechanics} \bibinfo{volume}{67},
  \bibinfo{pages}{1099--1114}.
\bibitem[{Born and Wolf(2013)}]{born2013principles}
\bibinfo{author}{Born, M.}, \bibinfo{author}{Wolf, E.}, \bibinfo{year}{2013}.
\newblock \bibinfo{title}{Principles of optics: electromagnetic theory of
  propagation, interference and diffraction of light}.
\newblock \bibinfo{publisher}{Elsevier}.
\bibitem[{Borrvall and Petersson(2003)}]{borrvall2003topology}
\bibinfo{author}{Borrvall, T.}, \bibinfo{author}{Petersson, J.},
  \bibinfo{year}{2003}.
\newblock \bibinfo{title}{Topology optimization of fluids in stokes flow}.
\newblock \bibinfo{journal}{International journal for numerical methods in
  fluids} \bibinfo{volume}{41}, \bibinfo{pages}{77--107}.
\bibitem[{Bourdin(2001)}]{bourdin2001filters}
\bibinfo{author}{Bourdin, B.}, \bibinfo{year}{2001}.
\newblock \bibinfo{title}{Filters in topology optimization}.
\newblock \bibinfo{journal}{International journal for numerical methods in
  engineering} \bibinfo{volume}{50}, \bibinfo{pages}{2143--2158}.
\bibitem[{Boyd and Vandenberghe(2004)}]{boyd2004convex}
\bibinfo{author}{Boyd, S.P.}, \bibinfo{author}{Vandenberghe, L.},
  \bibinfo{year}{2004}.
\newblock \bibinfo{title}{Convex optimization}.
\newblock \bibinfo{publisher}{Cambridge university press}.
\bibitem[{Bruns and Tortorelli(2001)}]{bruns2001topology}
\bibinfo{author}{Bruns, T.E.}, \bibinfo{author}{Tortorelli, D.A.},
  \bibinfo{year}{2001}.
\newblock \bibinfo{title}{Topology optimization of non-linear elastic
  structures and compliant mechanisms}.
\newblock \bibinfo{journal}{Computer methods in applied mechanics and
  engineering} \bibinfo{volume}{190}, \bibinfo{pages}{3443--3459}.
\bibitem[{Christiansen(2023)}]{christiansen2023inverse}
\bibinfo{author}{Christiansen, R.E.}, \bibinfo{year}{2023}.
\newblock \bibinfo{title}{Inverse design of optical mode converters by topology
  optimization: tutorial}.
\newblock \bibinfo{journal}{Journal of Optics} \bibinfo{volume}{25},
  \bibinfo{pages}{083501}.
\bibitem[{Christiansen et~al.(2019)Christiansen, Wang and
  Sigmund}]{christiansen2019topological}
\bibinfo{author}{Christiansen, R.E.}, \bibinfo{author}{Wang, F.},
  \bibinfo{author}{Sigmund, O.}, \bibinfo{year}{2019}.
\newblock \bibinfo{title}{Topological insulators by topology optimization}.
\newblock \bibinfo{journal}{Physical Review Letters} \bibinfo{volume}{122},
  \bibinfo{pages}{234502}.
\bibitem[{Dilgen and Aage(2024)}]{dilgen2024topology}
\bibinfo{author}{Dilgen, C.B.}, \bibinfo{author}{Aage, N.},
  \bibinfo{year}{2024}.
\newblock \bibinfo{title}{Topology optimization of transient vibroacoustic
  problems for broadband filter design using cut elements}.
\newblock \bibinfo{journal}{Finite Elements in Analysis and Design}
  \bibinfo{volume}{234}, \bibinfo{pages}{104123}.
\bibitem[{D{\"u}hring et~al.(2008)D{\"u}hring, Jensen and
  Sigmund}]{duhring2008acoustic}
\bibinfo{author}{D{\"u}hring, M.B.}, \bibinfo{author}{Jensen, J.S.},
  \bibinfo{author}{Sigmund, O.}, \bibinfo{year}{2008}.
\newblock \bibinfo{title}{Acoustic design by topology optimization}.
\newblock \bibinfo{journal}{Journal of sound and vibration}
  \bibinfo{volume}{317}, \bibinfo{pages}{557--575}.
\bibitem[{Felippa(2004)}]{felippa2004introduction}
\bibinfo{author}{Felippa, C.A.}, \bibinfo{year}{2004}.
\newblock \bibinfo{title}{Introduction to finite element methods}.
\newblock \bibinfo{journal}{University of Colorado} \bibinfo{volume}{885}.
\bibitem[{Gao et~al.(2020)Gao, Wang and Sigmund}]{gao2020systematic}
\bibinfo{author}{Gao, W.}, \bibinfo{author}{Wang, F.},
  \bibinfo{author}{Sigmund, O.}, \bibinfo{year}{2020}.
\newblock \bibinfo{title}{Systematic design of high-q prestressed micro
  membrane resonators}.
\newblock \bibinfo{journal}{Computer Methods in Applied Mechanics and
  Engineering} \bibinfo{volume}{361}, \bibinfo{pages}{112692}.
\bibitem[{Gertler et~al.(2023)Gertler, Kuang, Christie and
  Miller}]{gertler2023many}
\bibinfo{author}{Gertler, S.}, \bibinfo{author}{Kuang, Z.},
  \bibinfo{author}{Christie, C.}, \bibinfo{author}{Miller, O.D.},
  \bibinfo{year}{2023}.
\newblock \bibinfo{title}{Many physical design problems are sparse {QCQPs}}.
\newblock \bibinfo{journal}{arXiv preprint arXiv:2303.17691} .
\bibitem[{Grant and Boyd(2014)}]{grant2014cvx}
\bibinfo{author}{Grant, M.}, \bibinfo{author}{Boyd, S.}, \bibinfo{year}{2014}.
\newblock \bibinfo{title}{{CVX}: Matlab software for disciplined convex
  programming, version 2.1}.
\bibitem[{Grant and Boyd(2008)}]{grant2008graph}
\bibinfo{author}{Grant, M.C.}, \bibinfo{author}{Boyd, S.P.},
  \bibinfo{year}{2008}.
\newblock \bibinfo{title}{Graph implementations for nonsmooth convex programs},
  in: \bibinfo{booktitle}{Recent advances in learning and control},
  \bibinfo{organization}{Springer}. pp. \bibinfo{pages}{95--110}.
\bibitem[{Guest and Pr{\'e}vost(2006)}]{guest2006optimizing}
\bibinfo{author}{Guest, J.K.}, \bibinfo{author}{Pr{\'e}vost, J.H.},
  \bibinfo{year}{2006}.
\newblock \bibinfo{title}{Optimizing multifunctional materials: design of
  microstructures for maximized stiffness and fluid permeability}.
\newblock \bibinfo{journal}{International Journal of Solids and Structures}
  \bibinfo{volume}{43}, \bibinfo{pages}{7028--7047}.
\bibitem[{Hashin and Shtrikman(1963)}]{hashin1963variational}
\bibinfo{author}{Hashin, Z.}, \bibinfo{author}{Shtrikman, S.},
  \bibinfo{year}{1963}.
\newblock \bibinfo{title}{A variational approach to the theory of the elastic
  behaviour of multiphase materials}.
\newblock \bibinfo{journal}{Journal of the Mechanics and Physics of Solids}
  \bibinfo{volume}{11}, \bibinfo{pages}{127--140}.
\bibitem[{Jensen and Sigmund(2005)}]{jensen2005topology}
\bibinfo{author}{Jensen, J.S.}, \bibinfo{author}{Sigmund, O.},
  \bibinfo{year}{2005}.
\newblock \bibinfo{title}{Topology optimization of photonic crystal structures:
  a high-bandwidth low-loss t-junction waveguide}.
\newblock \bibinfo{journal}{JOSA B} \bibinfo{volume}{22},
  \bibinfo{pages}{1191--1198}.
\bibitem[{Jensen and Sigmund(2011)}]{jensen2011topology}
\bibinfo{author}{Jensen, J.S.}, \bibinfo{author}{Sigmund, O.},
  \bibinfo{year}{2011}.
\newblock \bibinfo{title}{Topology optimization for nano-photonics}.
\newblock \bibinfo{journal}{Laser \& Photonics Reviews} \bibinfo{volume}{5},
  \bibinfo{pages}{308--321}.
\bibitem[{Kim et~al.(2011)Kim, Kojima, Mevissen and
  Yamashita}]{kim2011exploiting}
\bibinfo{author}{Kim, S.}, \bibinfo{author}{Kojima, M.},
  \bibinfo{author}{Mevissen, M.}, \bibinfo{author}{Yamashita, M.},
  \bibinfo{year}{2011}.
\newblock \bibinfo{title}{Exploiting sparsity in linear and nonlinear matrix
  inequalities via positive semidefinite matrix completion}.
\newblock \bibinfo{journal}{Mathematical programming} \bibinfo{volume}{129},
  \bibinfo{pages}{33--68}.
\bibitem[{Kuang(2023)}]{kuang2023fundamental}
\bibinfo{author}{Kuang, Z.}, \bibinfo{year}{2023}.
\newblock \bibinfo{title}{Fundamental Limits of Nanophotonic Design}.
\newblock \bibinfo{publisher}{Yale University}.
\bibitem[{Kuang and Miller(2020)}]{kuang2020computational}
\bibinfo{author}{Kuang, Z.}, \bibinfo{author}{Miller, O.D.},
  \bibinfo{year}{2020}.
\newblock \bibinfo{title}{Computational bounds to light--matter interactions
  via local conservation laws}.
\newblock \bibinfo{journal}{Physical Review Letters} \bibinfo{volume}{125},
  \bibinfo{pages}{263607}.
\bibitem[{Liu et~al.(2018)Liu, Hu, Zhu, Matusik and Sifakis}]{liu2018narrow}
\bibinfo{author}{Liu, H.}, \bibinfo{author}{Hu, Y.}, \bibinfo{author}{Zhu, B.},
  \bibinfo{author}{Matusik, W.}, \bibinfo{author}{Sifakis, E.},
  \bibinfo{year}{2018}.
\newblock \bibinfo{title}{Narrow-band topology optimization on a sparsely
  populated grid}.
\newblock \bibinfo{journal}{ACM Transactions on Graphics (TOG)}
  \bibinfo{volume}{37}, \bibinfo{pages}{1--14}.
\bibitem[{Luo et~al.(2010)Luo, Ma, So, Ye and Zhang}]{luo2010semidefinite}
\bibinfo{author}{Luo, Z.Q.}, \bibinfo{author}{Ma, W.K.}, \bibinfo{author}{So,
  A.M.C.}, \bibinfo{author}{Ye, Y.}, \bibinfo{author}{Zhang, S.},
  \bibinfo{year}{2010}.
\newblock \bibinfo{title}{Semidefinite relaxation of quadratic optimization
  problems}.
\newblock \bibinfo{journal}{IEEE Signal Processing Magazine}
  \bibinfo{volume}{27}, \bibinfo{pages}{20--34}.
\bibitem[{Maute et~al.(1998)Maute, Schwarz and Ramm}]{maute1998adaptive}
\bibinfo{author}{Maute, K.}, \bibinfo{author}{Schwarz, S.},
  \bibinfo{author}{Ramm, E.}, \bibinfo{year}{1998}.
\newblock \bibinfo{title}{Adaptive topology optimization of elastoplastic
  structures}.
\newblock \bibinfo{journal}{Structural optimization} \bibinfo{volume}{15},
  \bibinfo{pages}{81--91}.
\bibitem[{Miller et~al.(2016)Miller, Polimeridis, Homer~Reid, Hsu, DeLacy,
  Joannopoulos, Solja{\v{c}}i{\'c} and Johnson}]{miller2016fundamental}
\bibinfo{author}{Miller, O.D.}, \bibinfo{author}{Polimeridis, A.G.},
  \bibinfo{author}{Homer~Reid, M.}, \bibinfo{author}{Hsu, C.W.},
  \bibinfo{author}{DeLacy, B.G.}, \bibinfo{author}{Joannopoulos, J.D.},
  \bibinfo{author}{Solja{\v{c}}i{\'c}, M.}, \bibinfo{author}{Johnson, S.G.},
  \bibinfo{year}{2016}.
\newblock \bibinfo{title}{Fundamental limits to optical response in absorptive
  systems}.
\newblock \bibinfo{journal}{Optics express} \bibinfo{volume}{24},
  \bibinfo{pages}{3329--3364}.
\bibitem[{Molesky et~al.(2020)Molesky, Chao and
  Rodriguez}]{molesky2020hierarchical}
\bibinfo{author}{Molesky, S.}, \bibinfo{author}{Chao, P.},
  \bibinfo{author}{Rodriguez, A.W.}, \bibinfo{year}{2020}.
\newblock \bibinfo{title}{Hierarchical mean-field t operator bounds on
  electromagnetic scattering: Upper bounds on near-field radiative purcell
  enhancement}.
\newblock \bibinfo{journal}{Physical Review Research} \bibinfo{volume}{2},
  \bibinfo{pages}{043398}.
\bibitem[{{Mosek ApS}(2019)}]{aps2019mosek}
\bibinfo{author}{{Mosek ApS}}, \bibinfo{year}{2019}.
\newblock \bibinfo{title}{Mosek optimization toolbox for matlab}.
\newblock \bibinfo{journal}{User’s Guide and Reference Manual, Version}
  \bibinfo{volume}{4}.
\bibitem[{Rosen and Hashin(1970)}]{rosen1970effective}
\bibinfo{author}{Rosen, B.W.}, \bibinfo{author}{Hashin, Z.},
  \bibinfo{year}{1970}.
\newblock \bibinfo{title}{Effective thermal expansion coefficients and specific
  heats of composite materials}.
\newblock \bibinfo{journal}{International Journal of Engineering Science}
  \bibinfo{volume}{8}, \bibinfo{pages}{157--173}.
\bibitem[{Shim et~al.(2024)Shim, Kuang, Lin and Miller}]{shim2024fundamental}
\bibinfo{author}{Shim, H.}, \bibinfo{author}{Kuang, Z.}, \bibinfo{author}{Lin,
  Z.}, \bibinfo{author}{Miller, O.D.}, \bibinfo{year}{2024}.
\newblock \bibinfo{title}{Fundamental limits to multi-functional and tunable
  nanophotonic response}.
\newblock \bibinfo{journal}{Nanophotonics} \bibinfo{volume}{13},
  \bibinfo{pages}{2107--2116}.
\bibitem[{Shor(1987)}]{shor1987quadratic}
\bibinfo{author}{Shor, N.Z.}, \bibinfo{year}{1987}.
\newblock \bibinfo{title}{Quadratic optimization problems}.
\newblock \bibinfo{journal}{Soviet Journal of Computer and Systems Sciences}
  \bibinfo{volume}{25}, \bibinfo{pages}{1--11}.
\bibitem[{Sigmund(2011)}]{sigmund2011usefulness}
\bibinfo{author}{Sigmund, O.}, \bibinfo{year}{2011}.
\newblock \bibinfo{title}{On the usefulness of non-gradient approaches in
  topology optimization}.
\newblock \bibinfo{journal}{Structural and Multidisciplinary Optimization}
  \bibinfo{volume}{43}, \bibinfo{pages}{589--596}.
\bibitem[{Sigmund and Maute(2013)}]{sigmund2013topology}
\bibinfo{author}{Sigmund, O.}, \bibinfo{author}{Maute, K.},
  \bibinfo{year}{2013}.
\newblock \bibinfo{title}{Topology optimization approaches: A comparative
  review}.
\newblock \bibinfo{journal}{Structural and multidisciplinary optimization}
  \bibinfo{volume}{48}, \bibinfo{pages}{1031--1055}.
\bibitem[{Sigmund and Torquato(1997)}]{sigmund1997design}
\bibinfo{author}{Sigmund, O.}, \bibinfo{author}{Torquato, S.},
  \bibinfo{year}{1997}.
\newblock \bibinfo{title}{Design of materials with extreme thermal expansion
  using a three-phase topology optimization method}.
\newblock \bibinfo{journal}{Journal of the Mechanics and Physics of Solids}
  \bibinfo{volume}{45}, \bibinfo{pages}{1037--1067}.
\bibitem[{Stolpe and Svanberg(2001)}]{stolpe2001alternative}
\bibinfo{author}{Stolpe, M.}, \bibinfo{author}{Svanberg, K.},
  \bibinfo{year}{2001}.
\newblock \bibinfo{title}{An alternative interpolation scheme for minimum
  compliance topology optimization}.
\newblock \bibinfo{journal}{Structural and Multidisciplinary Optimization}
  \bibinfo{volume}{22}, \bibinfo{pages}{116--124}.
\bibitem[{Str{\"o}mberg and Klarbring(2010)}]{stromberg2010topology}
\bibinfo{author}{Str{\"o}mberg, N.}, \bibinfo{author}{Klarbring, A.},
  \bibinfo{year}{2010}.
\newblock \bibinfo{title}{Topology optimization of structures in unilateral
  contact}.
\newblock \bibinfo{journal}{Structural and Multidisciplinary Optimization}
  \bibinfo{volume}{41}, \bibinfo{pages}{57--64}.
\bibitem[{Svanberg(1987)}]{svanberg1987method}
\bibinfo{author}{Svanberg, K.}, \bibinfo{year}{1987}.
\newblock \bibinfo{title}{The method of moving asymptotes—a new method for
  structural optimization}.
\newblock \bibinfo{journal}{International journal for numerical methods in
  engineering} \bibinfo{volume}{24}, \bibinfo{pages}{359--373}.
\bibitem[{Tortorelli and Michaleris(1994)}]{tortorelli1994design}
\bibinfo{author}{Tortorelli, D.A.}, \bibinfo{author}{Michaleris, P.},
  \bibinfo{year}{1994}.
\newblock \bibinfo{title}{Design sensitivity analysis: overview and review}.
\newblock \bibinfo{journal}{Inverse problems in Engineering}
  \bibinfo{volume}{1}, \bibinfo{pages}{71--105}.
\bibitem[{Vandenberghe and Boyd(1996)}]{vandenberghe1996semidefinite}
\bibinfo{author}{Vandenberghe, L.}, \bibinfo{author}{Boyd, S.},
  \bibinfo{year}{1996}.
\newblock \bibinfo{title}{Semidefinite programming}.
\newblock \bibinfo{journal}{SIAM review} \bibinfo{volume}{38},
  \bibinfo{pages}{49--95}.
\bibitem[{Wallin et~al.(2016)Wallin, J{\"o}nsson and
  Wingren}]{wallin2016topology}
\bibinfo{author}{Wallin, M.}, \bibinfo{author}{J{\"o}nsson, V.},
  \bibinfo{author}{Wingren, E.}, \bibinfo{year}{2016}.
\newblock \bibinfo{title}{Topology optimization based on finite strain
  plasticity}.
\newblock \bibinfo{journal}{Structural and multidisciplinary optimization}
  \bibinfo{volume}{54}, \bibinfo{pages}{783--793}.
\bibitem[{Wang et~al.(2018)Wang, Christiansen, Yu, M{\o}rk and
  Sigmund}]{wang2018maximizing}
\bibinfo{author}{Wang, F.}, \bibinfo{author}{Christiansen, R.E.},
  \bibinfo{author}{Yu, Y.}, \bibinfo{author}{M{\o}rk, J.},
  \bibinfo{author}{Sigmund, O.}, \bibinfo{year}{2018}.
\newblock \bibinfo{title}{Maximizing the quality factor to mode volume ratio
  for ultra-small photonic crystal cavities}.
\newblock \bibinfo{journal}{Applied Physics Letters} \bibinfo{volume}{113}.
\bibitem[{Wang et~al.(2014)Wang, Lazarov, Sigmund and
  Jensen}]{wang2014interpolation}
\bibinfo{author}{Wang, F.}, \bibinfo{author}{Lazarov, B.S.},
  \bibinfo{author}{Sigmund, O.}, \bibinfo{author}{Jensen, J.S.},
  \bibinfo{year}{2014}.
\newblock \bibinfo{title}{Interpolation scheme for fictitious domain techniques
  and topology optimization of finite strain elastic problems}.
\newblock \bibinfo{journal}{Computer Methods in Applied Mechanics and
  Engineering} \bibinfo{volume}{276}, \bibinfo{pages}{453--472}.
\bibitem[{Wang and Sigmund(2023)}]{wang2023architecting}
\bibinfo{author}{Wang, F.}, \bibinfo{author}{Sigmund, O.},
  \bibinfo{year}{2023}.
\newblock \bibinfo{title}{Architecting materials for extremal stiffness, yield,
  and buckling strength}.
\newblock \bibinfo{journal}{Programmable Materials} \bibinfo{volume}{1},
  \bibinfo{pages}{e6}.
\bibitem[{Yurtsever et~al.(2021)Yurtsever, Tropp, Fercoq, Udell and
  Cevher}]{yurtsever2021scalable}
\bibinfo{author}{Yurtsever, A.}, \bibinfo{author}{Tropp, J.A.},
  \bibinfo{author}{Fercoq, O.}, \bibinfo{author}{Udell, M.},
  \bibinfo{author}{Cevher, V.}, \bibinfo{year}{2021}.
\newblock \bibinfo{title}{Scalable semidefinite programming}.
\newblock \bibinfo{journal}{SIAM Journal on Mathematics of Data Science}
  \bibinfo{volume}{3}, \bibinfo{pages}{171--200}.

\end{thebibliography}

\section*{Appendix}
\subsection*{Passive regions}
In some topology optimization problems parts of the analysis domain might be
exempt from the design domain, i.e passive design regions can exist. Let the
design variables inside the passive regions be collected into the vector
$\bm{\theta}_c$, whereas the free design variables are collected in
$\bm{\theta}_f$. Using these index sets, we can decompose our linear system of
equations as
\begin{equation}
	\begin{bmatrix}
		\bm{A}_{cc} & \bm{A}_{cf} \\[10pt]
		\bm{A}_{fc} & \bm{A}_{ff}
	\end{bmatrix}\begin{bmatrix}
		\bm{z}_c \\[10pt]
		\bm{z}_f
	\end{bmatrix}
	= \begin{bmatrix}
		\bm{b}_c \\[10pt]
		\bm{b}_f
	\end{bmatrix}.
	\label{Decomposition}
\end{equation}
From the above we can solve for $\bm{z}_c$
\begin{equation}
	\bm{z}_{c} = \bm{A}_{cc}^{-1}\left(\bm{b}_c - \bm{A}_{cf}\bm{z}_f\right),
	\label{Schur1}
\end{equation}
and insert this in the equation for $\bm{z}_f$, which renders
\begin{equation}
	\bm{A}_{fc}\bm{A}_{cc}^{-1}\left(\bm{b}_c - \bm{A}_{cf}\bm{z}_f\right)  + \bm{A}_{ff}\bm{z}_{f} = \bm{b}_f \Leftrightarrow \left(\bm{A}_{ff} - \bm{A}_{fc}\bm{A}_{cc}^{-1}\bm{A}_{cf}\right)\bm{z}_f = \bm{b}_{f} - \bm{A}_{fc}\bm{A}_{cc}^{-1} \bm{b}_c.
	\label{Schur}
\end{equation}
By introducing $\tilde{\bm{A}} = \bm{A}_{ff} -
	\bm{A}_{fc}\bm{A}_{cc}^{-1}\bm{A}_{cf}$ and $\tilde{\bm{b}} = \bm{b}_{f} -
	\bm{A}_{fc}\bm{A}_{cc}^{-1} \bm{b}_c$, we see that the above takes the standard
form $\tilde{\bm{A}}\bm{z}_f = \tilde{\bm{b}}$, i.e. $\tilde{\bm{A}}$ and
$\tilde{\bm{b}}$ can replace ${\bm{A}}$ and ${\bm{b}}$ in \eqref{expandEqui1}.

\subsection*{FE derivations in the mode converter problem}
The weak form of \eqref{maxwellTM} over the mode converter domain $\Omega$ is
\begin{equation}
	\int_{\Omega}(\bm{\nabla}\delta E_z)^T\bm{\nabla}E_z \, dV - \int_{\Omega}k^2\varepsilon_r \delta E_z E_z \, dV - \int_{\partial\Omega} \delta E_z (\bm{\nabla}E_z)\cdot \bm{n} \, dS= 0.
	\label{weakform}
\end{equation}
We apply first-order absorbing boundary conditions on $\partial\Omega_{abs}\subset\partial\Omega$, such that
\begin{equation}
	\int_{\partial\Omega_{abs}} \delta E_z (\bm{\nabla}E_z)\cdot \bm{n} \, dS = - \int_{\partial\Omega_{abs}} ik\delta E_z E_z \, dS.
	\label{absorbingBC}
\end{equation}
On $\partial\Omega_{in}$ and $\partial\Omega_{out}$ we excite the input mode and extract the output mode, respectively. These modes are computed using the assumption that $E_z = E_z(x,y) = \tilde{E}(y)e^{-i\beta x}$. Inserting this in
\eqref{maxwellTM} renders
\begin{equation}
	\frac{d^2\tilde{E}}{d y^2} + (-\beta^2 + k^2\varepsilon_r) \tilde{E} = 0.
	\label{eigProb1}
\end{equation}
The weak form of the above reads
\begin{equation}
	\int_{\partial\Omega_{in/out}}\frac{d\delta\tilde{E}}{dy}\frac{d\tilde{E}}{dy} \, dy - \int_{\partial\Omega_{in/out}}k^2\varepsilon_r\delta\tilde{E}\tilde{E} \, dy - \left[\delta\tilde{E}\frac{d\tilde{E}}{dy} \right]_0^H = -\int_{\partial\Omega_{in/out}} \beta^2\delta\tilde{E}\tilde{E} \, dy.
	\label{eigProb2}
\end{equation}
Assuming perfect conductors, i.e. vanishing electric fields $E_z(y=0) = E_z(y=H)=0$, renders $\left[\delta\tilde{E}\frac{d\tilde{E}}{dy} \right]_0^H = 0$. After solving the generalized eigenvalue problem in \eqref{eigProb2} for the eigenpairs $(-\beta_{in}^2,\tilde{E}_{in})$ and $(-\beta_{out}^2,\tilde{E}_{out})$, using e.g. first order 1D line elements, we obtain $E_z^{eig} = \tilde{E}_{in}(y)e^{-i\beta_{in} x}$. In this way, the final boundary conditions in \eqref{weakform} are
\begin{equation}
	\begin{array}{ll}
		\ds\int_{\partial\Omega_{in}} \delta E_z (\bm{\nabla}E_z)\cdot \bm{n} \, dS = - \int_{\partial\Omega_{in}} i\beta_{in}\delta E_z E_z \, dS + \int_{\partial\Omega_{in}} 2i\beta_{in}\delta E_z E_z^{eig} \, dS , \\[10pt]
		\ds\int_{\partial\Omega_{out}} \delta E_z (\bm{\nabla}E_z)\cdot \bm{n} \, dS = - \int_{\partial\Omega_{out}} i\beta_{out}\delta E_z E_z \, dS
	\end{array}
	\label{eigBC}
\end{equation}
Putting \eqref{absorbingBC} and \eqref{eigBC} into \eqref{weakform} we obtain
\begin{equation}
	\begin{array}{ll}
		\ds\int_{\Omega}(\bm{\nabla}\delta E_z)^T\bm{\nabla}E_z \, dV  + \int_{\partial\Omega_{abs}} ik\delta E_z E_z \, dS - \int_{\Omega}k^2\varepsilon_r \delta E_z E_z \, dV \\[10pt]
		\ds+ \int_{\partial\Omega_{in}} i\beta_{in}\delta E_z E_z \, dS + \int_{\partial\Omega_{out}} i\beta_{out}\delta E_z E_z \, dS = \int_{\partial\Omega_{in}} 2i\beta_{in}\delta E_z E_z^{eig} \, dS.
	\end{array}
	\label{weakformFin}
\end{equation}
A finite element discretization of the above renders
\begin{equation}
	\bm{A}(\bm{\theta})\bm{z} = (\bm{K} + i\bm{D}(k) - k^2\bm{M}(\theta))\bm{z} = \bm{b},
	\label{FEmode}
\end{equation}
where
\begin{equation}
	\begin{array}{ll}
		\ds\bm{K} = \int_{\Omega}\bm{B}^T\bm{B} \, dV, \quad \bm{M}(\theta) = \text{diag}\left(\frac{1}{2}(\bm{\theta}^T(\varepsilon_r-1) + (\varepsilon_r+1)) \bm{M}^o\right), \quad \bm{b} = \int_{\partial\Omega_{in}} 2i\beta_{in}\bm{N}^T E_z^{eig} \, dS, \\[10pt]
		\ds\bm{D}(k) = \int_{\partial\Omega_{abs}} k\bm{N}^T\bm{N} \, dS
		\ds+ \int_{\partial\Omega_{in}} \beta_{in}\bm{N}^T\bm{N} \, dS + \int_{\partial\Omega_{out}} \beta_{out}\bm{N}^T\bm{N} \, dS.
	\end{array}
	\label{modeMatrices}
\end{equation}
In the above, $\bm{B}$ contains the gradients of the shape functions $\bm{N}$.

\subsection*{Mass-lumping scheme}
$\bm{M}^o$ is the HRZ lumped mass matrix, i.e. it is diagonal. The lumping procedure is based on \citet{felippa2004introduction}, and proceeds as follows:
\begin{itemize}
	\item[1.] Sum diagonal entries pertaining to translational (i.e. excluding  possibly rotational) degrees-of-freedoms of the consistent mass matrix, excluding any scaling factor.
	\item[2.] Divide each entry of the diagonal by this sum.
	\item[3.] Scale each entry by the total mass.
	\item[4.] (If applicable) Set each diagonal entry pertaining to the rotational degrees-of-freedom to zero.
\end{itemize}

\end{document}